\documentclass[prb,amsmath,amssymb,twocolumn, superscriptaddress,showpacs]{revtex4}

\usepackage{graphicx}
\usepackage{dcolumn}
\usepackage{bm}
\usepackage{braket}
\usepackage{relsize}

\begin{document}


\title{Dynamics of a mesoscopic nuclear spin ensemble interacting with an optically driven electron spin}

\author{M. J. Stanley}
\affiliation{Cavendish Laboratory, University of Cambridge, JJ Thomson Avenue, Cambridge CB3 0HE, United Kingdom}
\author{C. Matthiesen}%
\affiliation{Cavendish Laboratory, University of Cambridge, JJ Thomson Avenue, Cambridge CB3 0HE, United Kingdom}
\author{J. Hansom}%
\affiliation{Cavendish Laboratory, University of Cambridge, JJ Thomson Avenue, Cambridge CB3 0HE, United Kingdom}
\author{C. Le Gall}%
\affiliation{Cavendish Laboratory, University of Cambridge, JJ Thomson Avenue, Cambridge CB3 0HE, United Kingdom}
\author{C. H. H. Schulte}%
\affiliation{Cavendish Laboratory, University of Cambridge, JJ Thomson Avenue, Cambridge CB3 0HE, United Kingdom}
\author{E. Clarke}%
\affiliation{EPSRC National Centre for III-V Technologies, University of Sheffield, Sheffield, S1 3JD, UK}
\author{M. Atat{\"u}re}%
\email[Electronic address: ]{ma424@cam.ac.uk}
\affiliation{Cavendish Laboratory, University of Cambridge, JJ Thomson Avenue, Cambridge CB3 0HE, United Kingdom}

\date{\today}

\begin{abstract}
The ability to discriminate between simultaneously occurring noise sources in the local environment of semiconductor InGaAs quantum dots, such as electric and magnetic field fluctuations, is key to understanding their respective dynamics and their effect on quantum dot coherence properties. We present a discriminatory approach to all-optical sensing based on two-color resonance fluorescence of a quantum dot charged with a single electron. Our measurements show that local magnetic field fluctuations due to nuclear spins in the absence of an external magnetic field are described by two correlation times, both in the microsecond regime. The nuclear spin bath dynamics show a strong dependence on the strength of resonant probing, with correlation times increasing by a factor of four as the optical transition is saturated. We interpret the behavior as motional averaging of both the Knight field of the resident electron spin and the hyperfine-mediated nuclear spin-spin interaction due to optically-induced electron spin flips.

\end{abstract}

\maketitle

\section{\label{sec:I}Introduction}

Semiconductor quantum dots (QDs) allow deterministic trapping and manipulation of single charge and spin carriers in a solid-state system\cite{gammon2002optical}. Carrier wavefunctions are spread over the $10^{4}$-$10^{5}$ atoms that define the QD, giving rise to a large oscillator strength and the permanent dipole moment of the excited state\cite{warburton2002giant}. Interactions of the QD ground and optically excited states with electric and magnetic fields are manifest in the Zeeman splitting of spin states and DC Stark shifts of transition energies\cite{warburton2002giant, bayer2002fine, finley2002fine}. While the sensitivity to ambient fields can be exploited for metrology applications, for instance electrometry\cite{vamivakas2011nanoscale}, or optomechanical coupling\cite{wilson2004laser, yeo2013strain, montinaro2014quantum}, QDs constantly sense fields arising from the interaction with uncontrolled charges of the environment and the QD's bath of nuclear spins. The resulting inhomogeneous dephasing of a confined spin and the reduction of photon quality are particularly detrimental to application in emergent quantum technologies, where QD spins and photons have shown promise as qubit candidates\cite{warburton2013single,lodahl2013interfacing}. Hence, with regards to applications there is great interest in identifying and characterizing environmental fluctuation processes.

The physics underlying the environment fluctuations can be traced back to a range of solid-state interactions of fundamental interest. In particular, the interaction of a single resident electron spin with the bath of $N\sim 10^{4}$-$10^{5}$ nuclear spins exposes a multitude of interesting effects\cite{abragam1998principles,urbaszek2013nuclear,chekhovich2013nuclear} that are inherent to the photophysics of QDs. The contact hyperfine interaction can be described as an effective magnetic field acting on the electron spin where the fluctuation magnitude scales as $1/\sqrt{N}$. This instance of the `central spin problem'
has been widely studied theoretically\cite{merkulov2002electron, khaetskii2002electron, coish2004hyperfine,cywinski2009electron,sinitsyn2012role} and the resulting electron spin relaxation is expected to comprise three components with distinct dynamics: electron spin precession in the effective magnetic field of the nuclei (Overhauser field), nuclear spin precession in the effective magnetic field of the electron spin (Knight field), and nuclear spin dipolar interactions. The inhomogeneous electron spin dephasing occurring over a few nanoseconds as a consequence of precession in the slowly changing nuclear Overhauser field is well understood and measured\cite{johnson2005triplet, xu2008coherent, press2010ultrafast}. Surprisingly, first experimental data on the timescales of the nuclear spin bath dynamics in QDs have only recently emerged, reporting in one case correlation times of 100 $\mathrm{\mu}$s (5.5 $\mathrm{\mu}$s) for a resonantly driven negatively charged (neutral) QD\cite{kuhlmann2013charge}, and in the other case nuclear coherence times of a few milliseconds for a neutral QD\cite{chekhovich2014quadrupolar}. The dynamics, assigned to nuclear dipolar coupling in both reports, were obtained in the absence of an external magnetic field in the former and at fields of a few Tesla in the latter case. Studying nuclear spin bath dynamics for a driven QD is complicated by the simultaneously occurring electric field fluctuations which mask the optical signatures of the nuclear bath evolution. Recently, the advantage of resonant excitation for sensitive measurements was demonstrated by Kuhlmann \textit{et al.}, who identified two features in the power spectrum of QD resonance fluorescence attributed to electric and magnetic field noise\cite{kuhlmann2013charge}. However, a reliable method to isolate the effects of nuclear spin fluctuations, which would allow a direct study of their dynamics, is still missing.

In this work we introduce resonance fluorescence fluctuation spectroscopy as a general and highly sensitive approach to sensing the local environment of a QD. The resonant spectroscopy technique provides a means to quantify not only dynamics and magnitudes of electric field fluctuations, but also allows a direct study of nuclear spin dynamics. The paper is organized as follows: In Sec. II we discuss how electric and magnetic field fluctuations affect the QD's optical properties, focusing on the inherent fluctuations of the solid-state environment and their distribution functions. The experimental method is introduced in Sec. III where we use the intensity autocorrelation function to characterize the fluctuations for a single QD. Taking advantage of the excitonic transition's linear response to electric fields we use two-color excitation to isolate noise in the resonance fluorescence of a negatively charged QD solely due to magnetic field fluctuations. Consequently, we unambiguously identify two timescales associated with nuclear spin dynamics. Both are shorter than the $\sim$ 100 $\mathrm{\mu}$s expected for nuclear spin bath relaxation via a dipolar interaction in bulk material, but longer than $\sim$ 100 ns, which is predicted for electron spin dephasing as consequence of the nuclei precessing in the Knight field\cite{merkulov2002electron}. In Sec.  IV we find a strong dependence of these timescales on the optical driving strength. We discuss the relevance of optically induced electron spin flips to nuclear spin dynamics, providing a tentative explanation for our observations. Finally, we extract the time-averaged magnitudes of both electric and magnetic field fluctuations for several QDs in Sec. V. We show that the time-averaged fluctuations are consistent with Gaussian electric and nuclear field distribution functions. The standard deviation of those distributions, together with the timescales, fully quantifies the QD's local environment.

\section{\label{sec:II}Environment Noise Sources}
\begin{figure}[tb]
\includegraphics[width=\columnwidth]{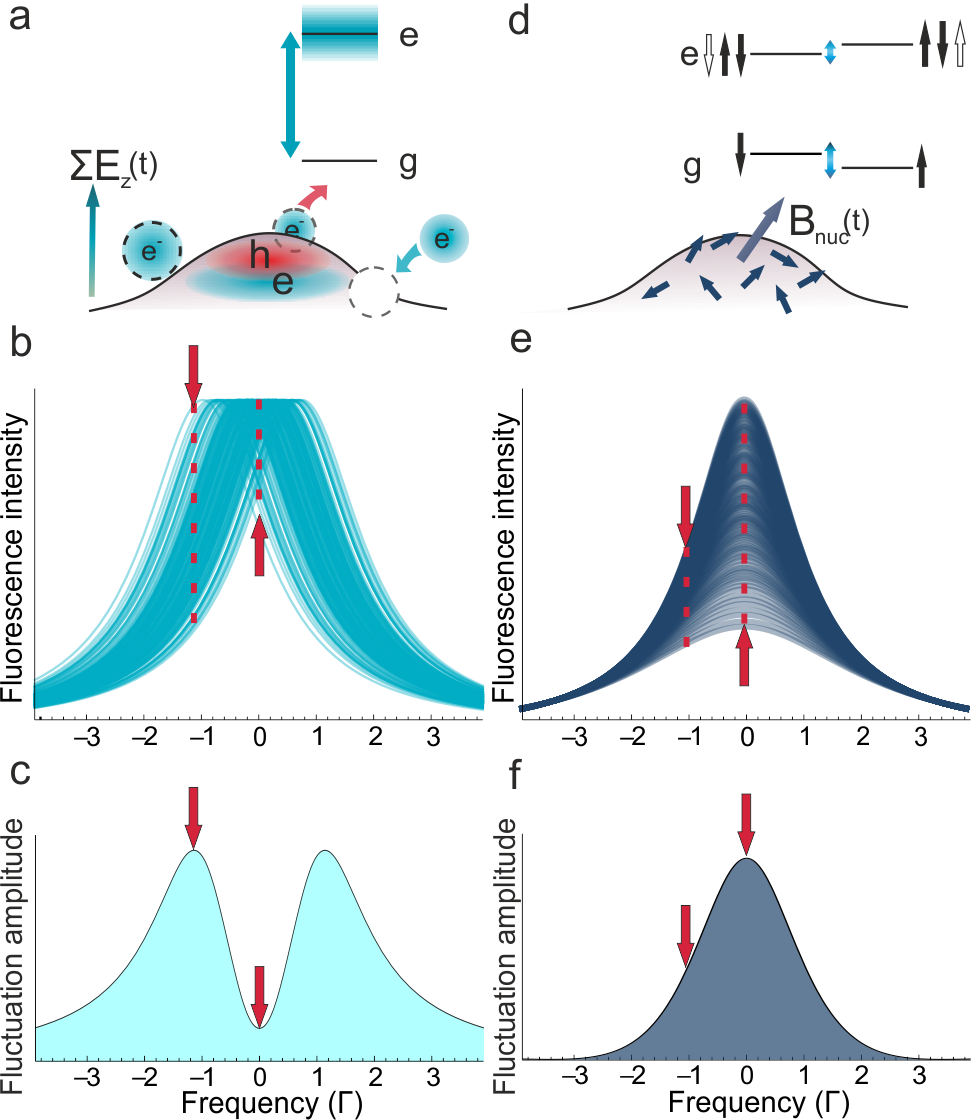}
\caption{\label{fig:1}(Color online)(a) A QD excited state energy level undergoes a Stark shift proportional to a change of the electric field component aligned with the dipole. Time-varying electric fields from local defects broaden the resonance of the optical transition. (b) QD absorption jitter for set of electric field values $E_{z}$ following a Gaussian probability distribution. The variance of the fluorescence intensity, indicated by dotted red lines, depends on the resonant laser detuning, here shown in units of the natural linewidth $\Gamma$. (c) The sensitivity to electric field noise is given by the intensity variance calculated from (b), divided by the square of the absorption. (d) The nuclear spin bath acts via the hyperfine interaction to produce an effective magnetic field (Overhauser field). Splitting of ground states and a weak splitting of the excited states gives rise to a four-level system where optical selection rules change with the Overhauser field. (e) QD absorption jitter for a set of Overhauser field values from a three dimensional Gaussian distribution. Intensity variance is indicated by dotted red lines for two detunings. (f) Resulting sensitivity to Overhauser field fluctuations calculated from the intensity variance in (e).}
\end{figure}

Figure ~\ref{fig:1}  illustrates the effect of a fluctuating environment on the intensity of the QD's fluorescence. The X$^{1-}$ transition serves as a fluctuation sensor for both electric (left column) and magnetic (right column) fields. In its ground state the QD contains a single electron and an additional electron-hole pair (exciton) is added in the excited state. In Fig. ~\ref{fig:1}(a) we consider local charge traps and impurities with fluctuating occupancy as sources of a noisy electric field. The large permanent dipole of the QD exciton renders the transition frequency sensitive to the component in the QD growth-direction of this field [$E_{z}(t)$], leading to a time-dependent linear Stark shift. The local electric field strength is reflected in the instantaneous resonance frequency of the QD transition. In the limit of many contributing electric field sources observed over a long time period a Gaussian distribution is a good description for the electric field probability distribution\cite{matthiesen2014full}. Figure~\ref{fig:1}(b) depicts the resonance fluorescence intensity `jitter' in the QD absorption lineshape for such a distribution function. A measurement of the absorption lineshape that is slow compared to the timescale of fluctuations would yield a Voigt profile in this case. The amplitude of resonance fluorescence fluctuations due to electric field noise corresponds to the variance of the fluorescence in the jitter plot and we highlight its detuning dependence here (see red arrows). In the bottom panel [Fig.~\ref{fig:1}(c)] the ratio of fluorescence variance to the squared fluorescence mean is plotted as a function of detuning, which represents the normalized fluorescence fluctuation amplitude.

The effect of the interaction with the nuclear spin bath is described in Fig. ~\ref{fig:1}(d): the nuclear spins of Indium (In), Gallium (Ga) and Arsenic (As) interact primarily with the electron spin through the contact hyperfine interaction. The cumulative effect of the hyperfine interaction with the nuclear spin at each lattice site can be described by a single magnetic field, the Overhauser field\cite{urbaszek2013nuclear}. Dynamics of this magnetic field modify the electronic energy levels [cf. four-level system in Fig. ~\ref{fig:1}(d)] and, accordingly, fluorescence rates under resonant excitation at fixed frequency. We calculate the absorption lineshape for a particular Overhauser field magnitude and orientation from the optical Bloch equations for the four-level system (see Appendix A). The absorption jitter plot for the fluctuating nuclear spin bath [Fig. ~\ref{fig:1} (e)] is obtained by sampling over an isotropic Gaussian Overhauser field distribution function\cite{urbaszek2013nuclear,hansom2014environment}. We note that the model predicts a time-averaged Lorentzian absorption lineshape, where the linewidth at saturation power is a factor of 1.5 larger than the power-broadened linewidth of an ideal two-level system. An Overhauser field distribution of $\sigma_{B}=25$ mT standard deviation and an excited state lifetime of $\text{T}_{1}=700$ ps is assumed in this example calculation (see Appendix A).

The calculated fluorescence fluctuation amplitudes displayed in Figs. ~\ref{fig:1} (c), (f) can be recovered directly in experiments as bunching amplitudes in the autocorrelation of the resonance fluorescence. Measurement techniques and results for resonance fluorescence fluctuation spectroscopy (RFFS) will be introduced in the following section. We note that under electric field variation, the fluorescence fluctuation amplitude is reduced on resonance in comparison to excitation at an intermediate detuning, where the amplitude peaks and then decays as detuning is increased. In contrast, variations in the Overhauser field produce the largest fluorescence fluctuations at zero detuning and the sensitivity is clearly reduced at finite detuning. We employ the contrasting detuning dependence, pointed out in Ref.~\onlinecite{kuhlmann2013charge} before, in the following section for a qualitative interpretation of fluctuation amplitudes and again in section V to obtain numerical values for electric and magnetic field noise.

\section{\label{sec:III}Resonance fluorescence fluctuation spectroscopy}
\begin{figure*}[tb]
\includegraphics[width=1.5\columnwidth]{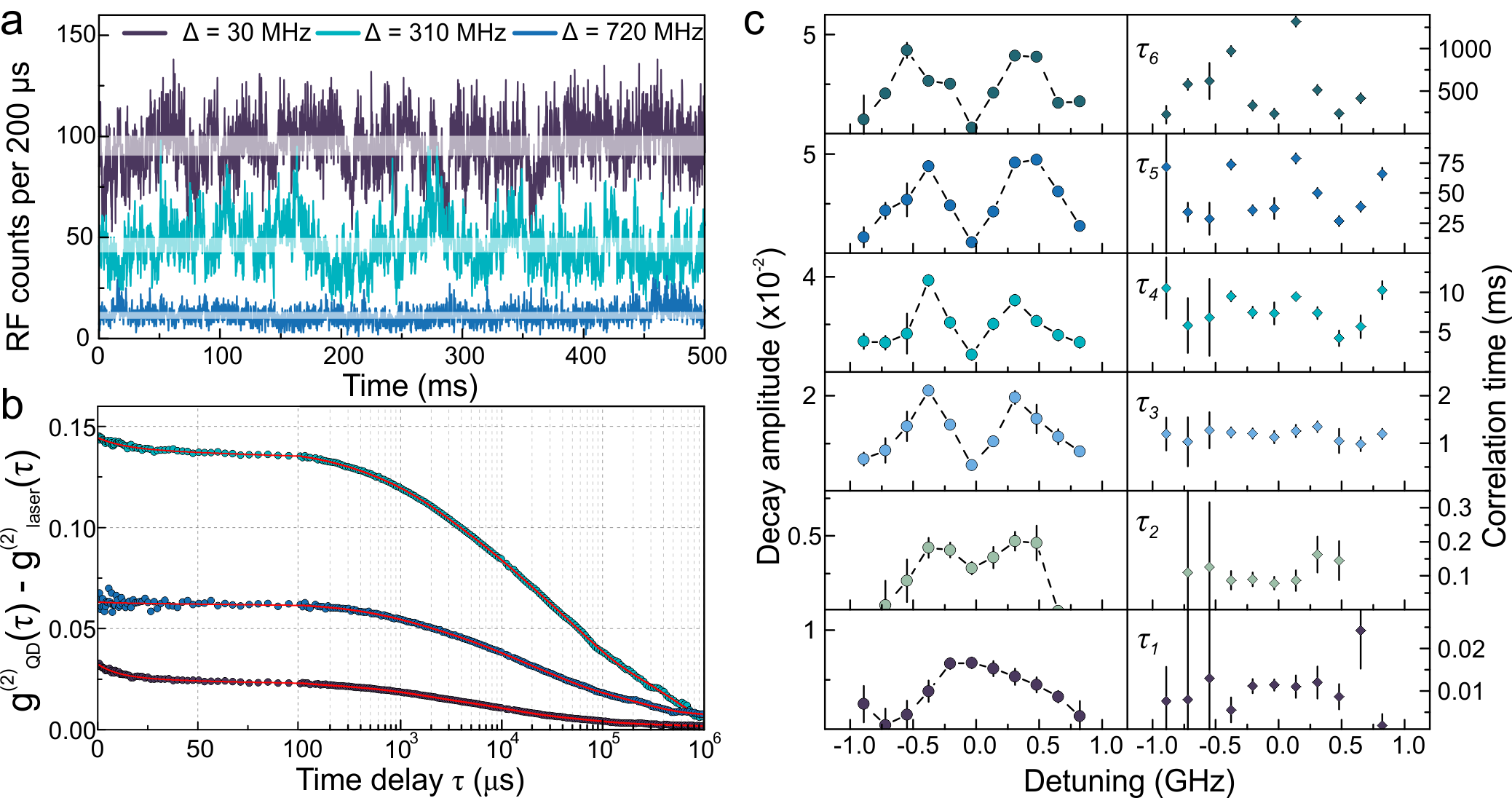}
\caption{\label{fig:2}(Color online) (a) Segments of resonance fluorescence time traces from QD A X$^{1-}$ driven below saturation, $s = 0.28$. APD counts are binned to 200 $\mu$s resolution here. White bars display the standard deviations expected from Poisson statistics (bar thickness) about the mean count rate. The respective excitation detuning is indicated in the legend. (b) Intensity autocorrelations calculated from time traces presented in (a), data as circles, fit as line. Bunching amplitudes vary significantly with detuning. (c) Fitting autocorrelations with multiple exponential decays reveals distinct decay timescales (right). Left: amplitudes of autocorrelation decays are strongly dependent on laser detuning.}
\end{figure*}
InGaAs QDs in a Schottky diode device are located in a liquid Helium bath cryostat at 4 K temperature and at 0 T external magnetic field. We use frequency and power-stabilized lasers to resonantly excite single QDs in continuous-wave mode and linear polarization. QD resonance fluorescence is collected by means of a confocal microscope in a dark-field configuration\cite{matthiesen2012subnatural} and detected by a single photon counting avalanche photodiode (APD). Photon arrival times are registered by a time-to-digital converter with a timing resolution of 81 ps and rebinned in post-processing. We present results for three QDs, labeled A, B and C in the main text and the appendices. 

Figure~\ref{fig:2} displays a set of RFFS measurements.  Three example photon detection time traces from QD A are displayed in Fig.~\ref{fig:2}(a) for excitation on resonance and detunings of $\Delta=310$ MHz and $\Delta=720$ MHz, where the natural linewidth of the transition, $\Gamma$, is 270 MHz in linear frequency.  The excitation power corresponds to a fifth of the saturation power, that is, $s=$ 0.2, where $s= 2\left(\mathit{\Omega}/\Gamma\right)^{2}$ and $\mathit{\Omega}$ is the Rabi frequency. The standard deviation expected due to Poissonian shot noise is indicated by the thickness of white semi-transparent stripes. To extract fluorescence dynamics over a wide range of timescales we use the intensity autocorrelation function g$^{(2)}(\tau)$, where the variable $\tau$ specifies the time delay between photodetections. Obtaining and analyzing the autocorrelation of a fluorescence signal is a well-known spectroscopy technique\cite{magde1972thermodynamic, krichevsky2002fluorescence}, for example used to quantify molecular diffusion dynamics\cite{lippitz2005statistical}. Here, we apply this technique to single QD resonance fluorescence, where fluctuations are instead due to the solid-state environment. In the autocorrelation function the shot noise limit corresponds to  g$^{(2)}(\tau) = 1$, while super-Poissonian correlation between photons will result in bunching, that is,  g$^{(2)}(\tau)$ \textgreater 1. Figure~\ref{fig:2}(b) displays the autocorrelations corresponding to the time traces of panel~\ref{fig:2}(a) which contain $\sim$ 10$^{7}$ time-tagged detection events for each time trace, corresponding to acquisition times of 100-200 seconds depending on the laser detuning. Systematic errors, mainly due to APD afterpulsing, were accounted for by taking- reference measurements of laser photon streams at comparable count rates, and subtraction of the corresponding autocorrelation from the QD resonance fluorescence autocorrelation (see Appendix B).

Fits of the experimental autocorrelations to a sum of exponential decays, shown as red lines in Fig.~\ref{fig:2}(b), reveal a set of distinct correlation times. In the case of telegraph noise a single exponential decay is expected\cite{machlup1954noise} and a set of correlation times indicates several fluctuation processes are present: for QD A we resolve six timescales ranging from about 10 $\mu$s to 1 s in the fit. Detailed data on timescales and amplitudes of the individual correlation decays are presented in Fig.~\ref{fig:2}(c). Amplitudes (left column) corresponding to correlation times (right column) of $\sim$ 1 ms and longer are clearly reduced on resonance. In contrast, the shortest correlation time amplitudes are maximal on resonance. We compare this detuning dependence with the discussion of noise amplitudes around Fig.~\ref{fig:1}, and discern that electric field fluctuations make the dominant contribution to noise on timescales of 1 ms and longer. We label these timescales $\tau_{3}-\tau_{6}$. In contrast, the detuning dependence of the $\tau_{1}$ process points to magnetic field fluctuations as source of noise. However, the large number of noise sources present for this QD can give rise to dependencies between fit parameters and make a direct identification challenging. The correlation amplitudes corresponding to $\tau_{2}$ ($\sim$ 100 $\mu$s) highlight the ambiguity in this approach: the detuning dependence does not fit into a single category, suggesting contributions from both noise sources. Similarly, we cannot exclude the presence of electric field noise in the fastest decay, at 10 $\mu$s, from this measurement while nuclear spin bath fluctuations could also be contributing to longer correlation decays. 

\begin{figure*}[tb]
\includegraphics[width=1.5\columnwidth]{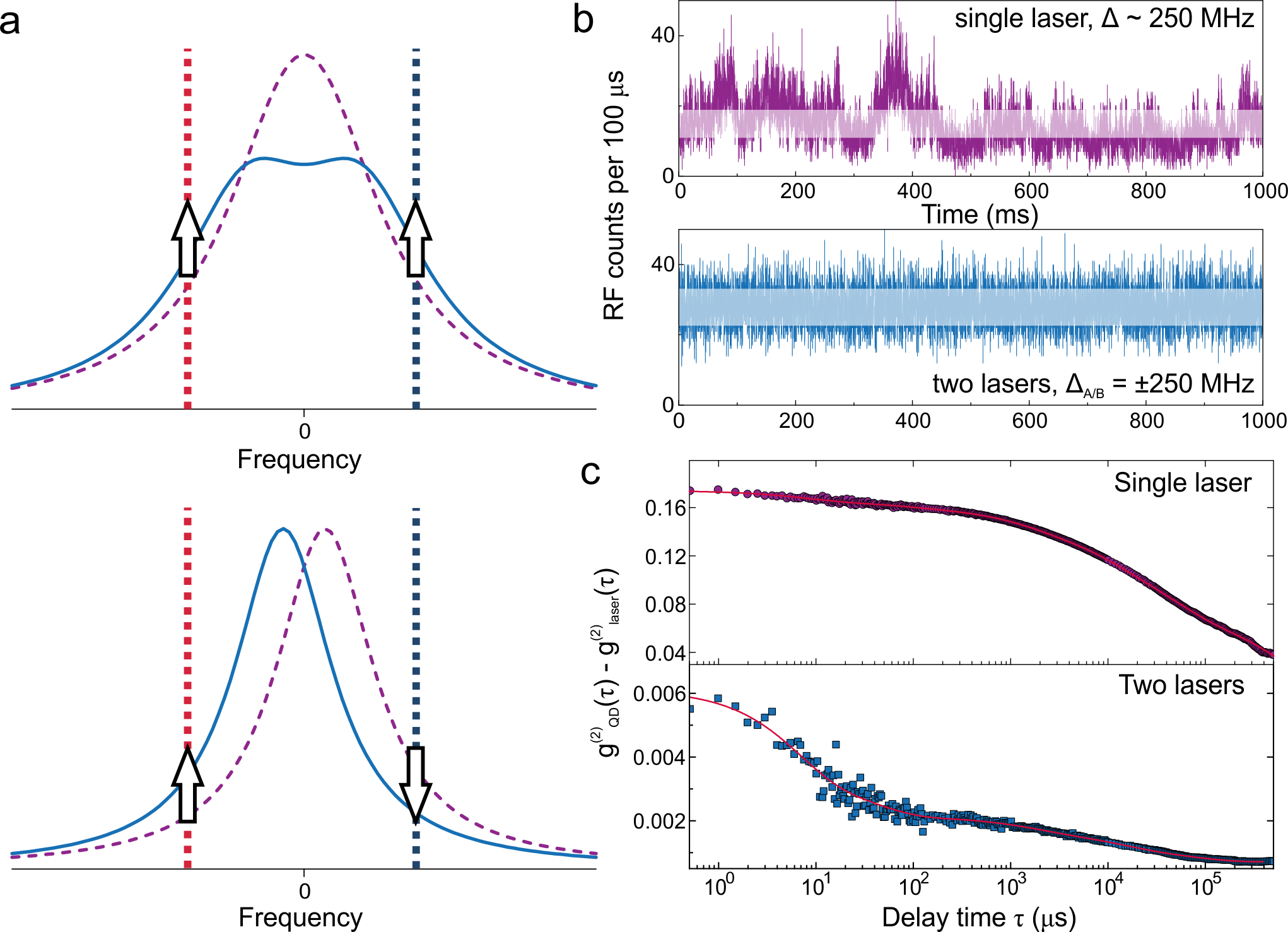}
\caption{\label{fig:3}(Color online) (a) Sketch of the noise balancing concept. Two-laser excitation at equal and opposite detuning renders the resonance fluorescence intensity insensitive to small linear shifts in QD resonance frequency caused by electric field fluctuations (bottom). Resonance fluorescence intensity noise due to Overhauser field changes is enhanced (top). (b) Upper panel: resonance fluorescence time trace segment for QD A X$^{1-}$ under excitation with a laser detuned by half a linewidth. Lower panel: resonance fluorescence from the same QD at identical total excitation power with two equally detuned lasers, enabling direct comparison. White bars indicate shot noise from the mean count rates; in the case of two lasers the fluctuations about this are much reduced in comparison to single laser excitation. (c) Autocorrelations of data from (b). Bunching amplitudes for time delays $>$ 100 $\mu$s are strongly suppressed for two-laser excitation. Bunching with characteristic decay times of 6 $\mu$s and 40 $\mu$s remains. The long timescale amplitude is reduced by about two orders of magnitude whilst noise on shorter timescales remains, consistent with a magnetic field origin.}
\end{figure*}

In order to discriminate the noise sources unambiguously we isolate magnetic field noise in the QD fluorescence using two-color excitation. The concept is illustrated in Fig.~\ref{fig:3}(a) where the effects of magnetic (top) and electric (bottom) field changes on fluorescence intensity are considered separately. At the top we consider the QD absorption lineshape for two values of the Overhauser field. An increase of the Overhauser field from a small to a large value splits the QD ground states and is reflected in a splitting of the absorption lineshape which changes from the dashed to the continuous curve. When two lasers of equal power drive the QD transition at equal and opposite detuning from resonance (indicated by vertical lines) the change in absorption is equal at both laser frequencies (cf. white arrows). In contrast, linear Stark shifts due to changes in the ambient electric field, illustrated in the bottom sketch, cause opposite changes in intensity of resonance fluorescence at each frequency. Figure~\ref{fig:3}(b) presents a resonance fluorescence time trace for excitation with a single laser (top) at a detuning $\Delta \sim$ 250 MHz, which yields half the fluorescence intensity compared to excitation on resonance. The bottom time trace corresponds to excitation with two lasers at detunings $\pm \Delta$. The total laser power incident on the sample is identical in both cases and corresponds to $s\approx 0.1$ . The autocorrelation [cf. Fig.~\ref{fig:3}(c)] for the two-laser excitation demonstrates a reduction of slow ($\tau >$ 1 ms) decay processes by up to two orders of magnitude in amplitude, while noise with short correlation times remains. The suppression of electric field-related noise in the fluorescence allows us to probe nuclear field fluctuations with greater clarity, revealing two distinct decays of $\tau_{\text{N}1}=6\mu$s and $\tau_{\text{N}2}= 40 \mu$s with similar amplitudes where the subscript $\text{N}$ specifies the origin as nuclear spin noise. The next fastest correlation decay happens on a 1.5 ms timescale and is reduced by a factor of 50 in comparison to single laser excitation, consistent with residual electric field fluctuations.
The correlation times measured here can be compared to the established model of nuclear spin dynamics in bulk GaAs and their effect on electron spin dephasing\cite{merkulov2002electron}: precession of nuclear spins in the Knight field of the electron is expected to cause electron spin relaxation on a timescale $\text{T}_{\text{K}}=1/\gamma_{\text{K}}$ of a few 100 ns while dipolar interactions between nuclear spins change the Overhauser field on a 100 $\mu$s timescale. However, strain in InGaAs QDs strongly modifies the dipolar interactions\cite{lai2006knight,chekhovich2014quadrupolar,welander2014influence} and experimental considerations such as the details of the sample structure and the impact of optical excitation must be taken into account. In the following we will associate the fastest timescale, $\tau_{\text{N}1}$ with an effective Knight field precession time, and the second timescale $\tau_{\text{N}2}$ with an effective nuclear spin-spin interaction time. 

\section{\label{sec:IV}Nuclear spin correlation times for a driven quantum dot}
Having established two timescales for magnetic field noise in the QD fluorescence we examine their dependence on external parameters. To obtain access to the detuning dependence we use single-laser excitation. The sensitivity to nuclear spin fluctuations is increased by selecting a different QD (QD B) on the same sample that has a smaller Stark coefficient, reducing the effect of electric field noise. It is also important to consider the excited state lifetime, as a short lifetime translates to a broad natural linewidth $\Gamma = (2 \pi \text{T}_{1})^{-1}$ and consequently a smaller sensitivity to noise in general. For QD A we measure T$_{1}$ = $(584\pm10)$ ps, however for QD B we measure T$_{1}$ = $(693\pm5)$ ps, yielding a greater overall sensitivity to noise. 
\begin{figure}[tb]
\includegraphics[width=\columnwidth]{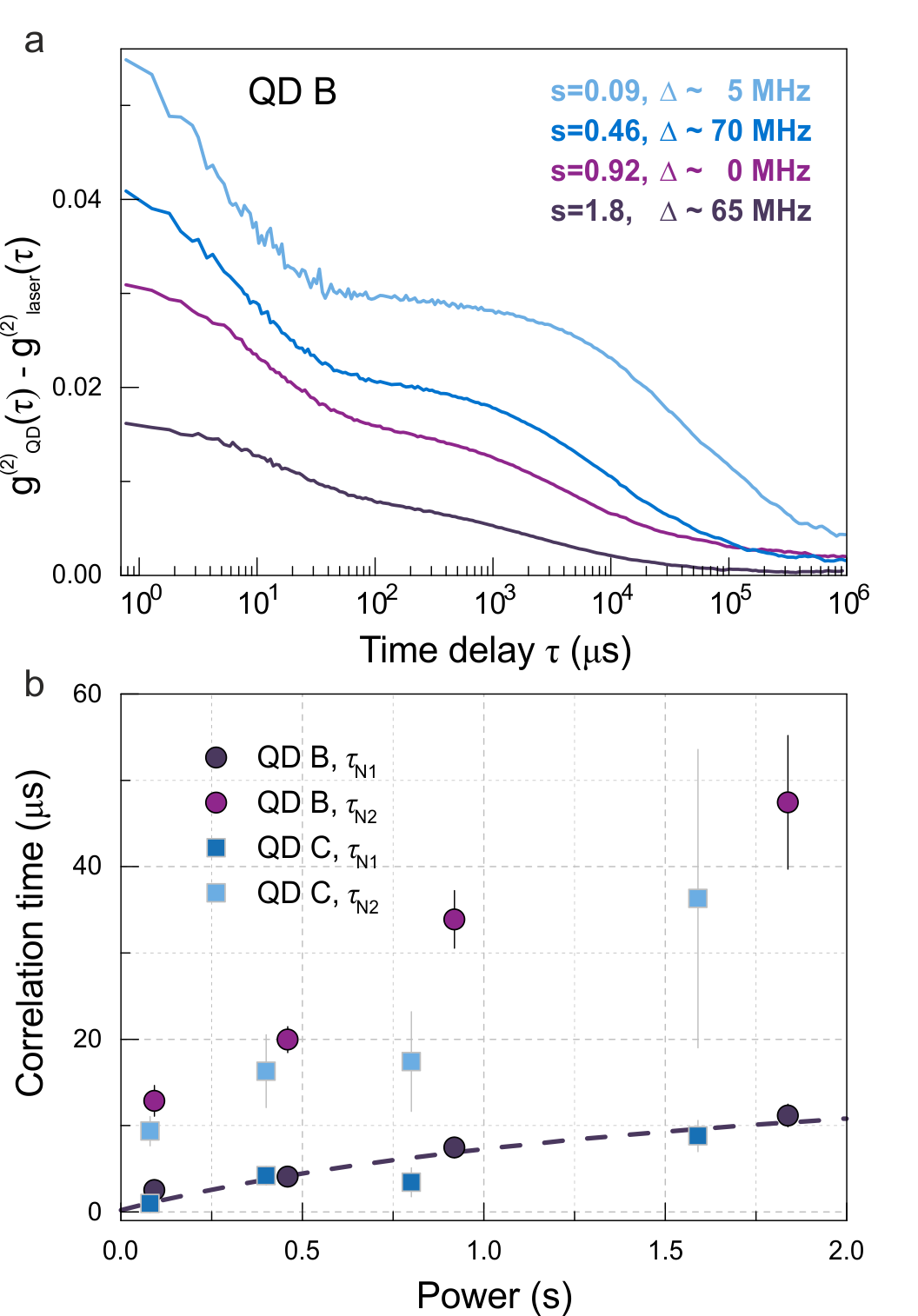}
\caption{\label{fig:4}(Color online) (a) Autocorrelations for QD B X$^{1-}$ for close to resonant ($\Delta$ $\sim$ 0 MHz) driving. The saturation parameter (indicated by the legend) is varied between $s=0.09$ and  $s=1.8$. (b) Correlation times of the nuclear spin bath, extracted from exponential fits to the data (not shown), show a strong dependence on the driving power. }
\end{figure}

Figure~\ref{fig:4}(a) displays four autocorrelations for excitation of QD B close to resonance. The excitation power is varied from $\textrm{s} = 0.1$ to $\textrm{s} = 2$. The bunching amplitude of the autocorrelation function decreases markedly as a consequence of power broadening. This effect is analogous to the dependence of noise sensitivity on the natural linewidth: as the excitation power is increased the inherent broadening of the absorption reduces sensitivity to all fluctuations and consequently noise amplitudes. More surprisingly, however, the dynamics at short time delays, which we identified to be due to nuclear spin fluctuations, slow down with increasing power. Figure~\ref{fig:4}(b) summarizes the power dependence of the fast timescales for QD B (dark filled circles). For reference we provide an additional set of data for QD C (light filled squares). Taking QD B data in particular, the correlation times increase from  $\tau_{\mathrm{N}1}$ = (2.5$\pm$0.5) $\mathrm{\mu}$s at $s = 0.09$ to  $\tau_{\mathrm{N}1}$ = (11$\pm$1) $\mu$s at $s = 1.8$. Similarly, $\tau_{\mathrm{N}2}$ increases from (13$\pm$2) $\mu$s to (47$\pm$8) $\mu$s in the same range. In fact, the ratio of correlation times is approximately constant in our measurements, giving $\tau_{\mathrm{N}2}/\tau_{\mathrm{N}1}\sim 4.5$ in this case. QD C shows qualitatively the same behavior. 

We first provide a tentative explanation for the $\tau_{\mathrm{N}1}$ dynamics here by identifying the primary mechanism for Overhauser field fluctuations as the precession of nuclear spins in the Knight field of the electron. A power dependence arises as a consequence of both the optically-induced spin flips of the electron and in addition the change of electron ground state population. First, we note that the Knight field is present while the QD is in the ground state. The field is negligible in the excited state as the electrons form a spin singlet and the heavy hole has a much weaker hyperfine interaction. Consequently, the electron spin's dephasing rate $\gamma_{\text{N}1}=1/\tau_{\text{N}1}$ should scale with the ground state population, and decrease in line with the optical saturation to half its maximum value at high probing power. Furthermore, the Knight field is affected by the electron spin lifetime. Electron spin flip rates $\gamma_{\text{sp}}$ comparable to, or larger than, the nuclear precession rate in the electron's Knight field result in a motional averaging and suppress the effect of the Knight field. Contributions to the electron spin flip rate in our experiments include spin-flip co-tunneling processes and optically induced spin flips. Spin-flip Raman transitions in the 4-level system (cf. Fig. \ref{fig:1}) are allowed for Overhauser field configurations with a component in the plane perpendicular to the growth axis. Spin pumping via this channel occurs on average after three optical cycles in the absence of an external field\cite{hansom2014environment} such that we obtain a cumulative spin flip rate:
\begin{equation}
\label{eq:sp}
\gamma_{\text{sp}}=\gamma_{\text{cot}}+\frac{s}{2\left(1+s\right)\text{T}_{1}}\times\frac{1}{3}\,.
\end{equation}
The co-tunneling spin-flip rate $\gamma_{\text{cot}}$ for this sample was measured using a protocol similar to the one detailed in Ref.~\onlinecite{lu2010direct} to be $\sim (100\mu\text{s})^{-1}$. The second term in Eq. \eqref{eq:sp} describes the rate of optically-induced spin flips. It gives spin flip times of tens of nanoseconds at low excitation power and below ten nanoseconds above saturation. Comparing the two contributions we note that the effect of co-tunneling is negligible in our device. We find that even for an excitation power corresponding to a tenth of the saturation power a significant motional averaging effect of the Knight field should take place, effectively prolonging electron spin dephasing due to nuclei precessing in the Knight field into the microsecond range. We may capture the dynamics in a phenomenological model of the competition of rates which is plotted in Fig. ~\ref{fig:4}(b) as a dashed line:
\begin{equation}
\label{eq:gamman1}
\gamma_{\text{N}1}\approx\frac{\gamma_{\text{K}}}{\gamma_{\text{K}}+\gamma_{\text{sp}}}\gamma_{\text{K}}\rho_{\text{g}},
\end{equation}
where $\gamma_{\text{K}}$ is the unperturbed electron spin dephasing rate arising from nuclear precession in the Knight field and $\rho_{\text{g}}$ is the QD ground state population. 

For a QD of $N$ nuclear spins the electron spin dephasing time due to Knight field precession scales as $\text{T}_{\text{K}}\sim\sqrt{N}\text{T}_{2,\text{e}}^{\star}$, where $\text{T}_{2,\text{e}}^{\star}\sim 1$ ns is the electron spin dephasing time in the Overhauser field\cite{merkulov2002electron}. For a QD of average size, $N\sim 5\times10^{4}$, we obtain $\text{T}_{\text{K}}=1/\gamma_{\text{K}}\sim200 \text{ns}$ which reproduces the power dependence we observe in the data. We note that electron spin relaxation on a timescale of 300 ns due to the Knight field has been reported very recently by Bechtold et al. \cite{bechtold2014hyperfine}

Concerning the origin of the correlation time $\tau_{\mathrm{N}2}$, we may exclude direct dipolar coupling of nuclear spins as the sole contributor because it is a local interaction that depends only weakly on dynamics of the electron spin state, or the QD ground state population. Instead, hyperfine-mediated indirect coupling of nuclear spins, which was shown to be an efficient mechanism for relaxation of dynamic nuclear spin polarization \cite{maletinsky2007dynamics,latta2011hyperfine} and electron spin dephasing\cite{deng2008electron,cywinski2009electron} is likely to be at the origin of the $\tau_{\mathrm{N}2}$ correlation. The interaction strength of this second-order process is at least equal to the dipolar interaction in bulk material and dominates dynamics in strained QD systems, where quadrupolar effects (and the Knight field) suppress dipolar coupling. Hyperfine-mediated nuclear spin interaction is dependent upon the electron spin state and as such will also be susceptible to motional averaging under electron spin flips in the same manner as nuclear spin precession in the electron Knight field. It remains an open question at this stage whether other (excitation-power dependent) interactions\cite{paget2008light} take part in nuclear spin dynamics at these timescales. The data indicate a correlation time $\tau_{\mathrm{N}2}\sim 10\;\mu$s in the absence of optical excitation. We find quantitatively similar behavior for different QDs on the same sample. We note that, in contrast to the nuclear spin dynamics, correlation times associated with electric field fluctuations display a speedup of about a factor two for the same increase of excitation power.

Our results clearly demonstrate a dependence of the dynamics of the nuclear spin bath on the strength of resonant optical excitation. Motional averaging due to spin flips of the resident electron provide qualitative agreement with our observations. The precise nuclear spin dynamics in other devices are expected to be sensitive to differences in the sample structure. Of particular importance is the size of the tunnel barrier separating the QD layer from the doped back contact which determines the electron spin-flip co-tunneling rate. Our experimental technique and the description with Eqs. \eqref{eq:sp}, \eqref{eq:gamman1} remain valid for different structures, but the relevance of the two contributions in Eq. \eqref{eq:sp} is modified. While the spin-flip co-tunneling timescale for our sample (35-nm barrier) is about 100 $\mu$s in the center of the one-electron stability plateau, a 25-nm barrier (compared to Ref.~\onlinecite{kuhlmann2013charge}) can result in $\text{T}_{\text{cot}}$ in the nanosecond regime, therefore dominating the rates in Eq. \eqref{eq:sp}. As a consequence of the fast spin recycling for narrow tunnel barriers, we expect the Knight field to be entirely absent, and electron-mediated nuclear spin interaction to be weak. In the limit of fast spin flips we expect to recover a nuclear bath fluctuation time governed by direct dipolar coupling.

\section{\label{sec:V}Quantifying electric and magnetic field fluctuations}
Magnetic and electric field correlation times for QDs in our device are well separated (up to 50 $\mu$s for nuclear spin bath fluctuations, beyond 1 ms for electric fields) so that electric field fluctuations can be considered frozen on the timescale of Overhauser and Knight field evolution. Here we employ this separation to quantify noise magnitudes using the model discussed in Fig.~\ref{fig:1}. We first calculate the time-averaged effect of a nuclear spin bath with isotropic distribution function on the excited state populations. Here, the sub-linewidth ground state splitting results in a broadened absorption lineshape (see Appendix A). Electric field fluctuations are then included as a Gaussian distribution of transition resonance frequencies. The electric field contribution to noise in the fluorescence is found directly as the ratio of the resulting variance to the square of the mean excited state population. Our experimental data contain several processes on different timescales associated with electric field fluctuations, however we are able to characterize the combined noise averaged over long measurement times with a single field distribution function\cite{matthiesen2014full}. In this case it is the sum of noise amplitudes that we are concerned with and therefore a non-Markovian model which treats dynamics on multiple timescales independently is not required.
\begin{figure}[tb]
\includegraphics[width=\columnwidth]{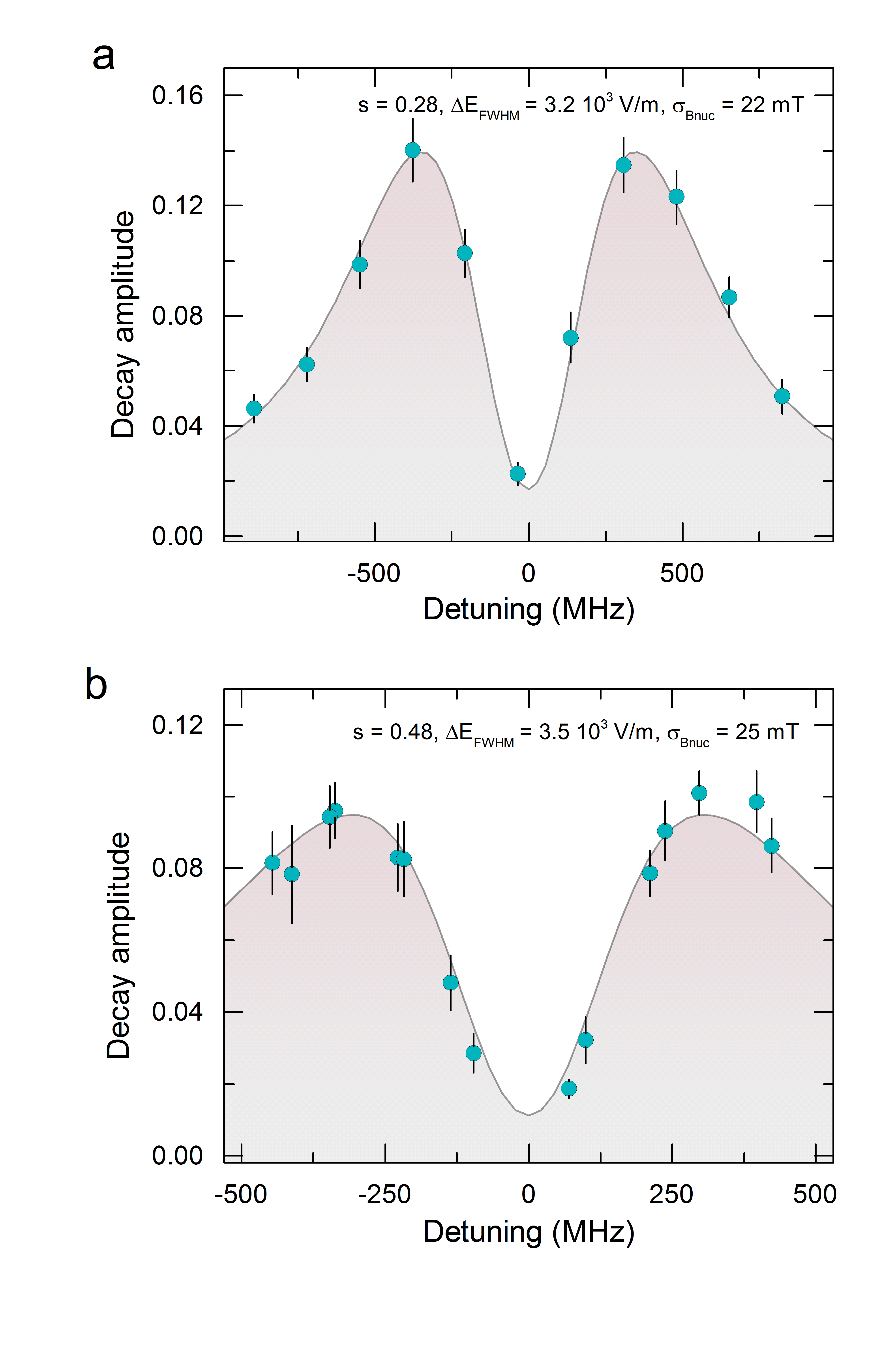}
\caption{\label{fig:5}(Color online) Example simulations of electric field noise, where measured amplitudes are displayed as blue circles. (a) Electric field fluctuation amplitudes for QD A. Here fast noise is masked by the large electric field noise contribution. (b) Comparison to simulation for QD B.}
\end{figure}

Taking the value of the measured autocorrelation function at a time delay where the dominant contributions due to the nuclear field fluctuations have decayed ($\tau$ $\sim$ 200 $\mu$s), we find the noise amplitude due to the electric field happening on all (longer) timescales. In Fig.~\ref{fig:5}(a) this fluctuation amplitude for QD A (data as circles) is fit using the time-averaged model (curve) where free parameters are $\Delta E_{\mathrm{FWHM}}$ representing the full-width-half-maximum of the electric field distribution function and the standard deviation of the Overhauser field distribution $\sigma_{\mathrm{B}}$. The simulation is in agreement with the data for an Overhauser field distribution with standard deviation $\sigma_{\mathrm{B}}=(22\pm 2)$ mT and a broadening of the optical transition by a Gaussian distribution with a FWHM of (205$\pm$7) MHz. Taking into account the measured Stark shift for this QD we arrive at an electric field fluctuation distribution with a FWHM of (3.2$\pm$0.1)10$^{3}$ V/m. For QD B we extract $\sigma_{\mathrm{B}}$ = (25$\pm$2) mT and  $\Delta E_{\mathrm{FWHM}} = (3.5\pm 0.2)10^{3}$ V/m, which corresponds to a transition frequency broadening due to electric field of  (168$\pm$11) MHz [Fig.~\ref{fig:5}(b)]. We note that whilst this model does not include the fluctuation processes of nuclear spins explicitly, it is possible to obtain a characteristic Overhauser field distribution through its necessary impact on the underlying absorption lineshape. The model is applicable for low to moderate QD excitation. In this regime we find the extracted Overhauser field distributions to be unaffected by excitation power.

The standard deviation of the Overhauser field distribution we extract from sets of autocorrelations are consistent between QDs and agree with values reported in the literature inferred through other techniques\cite{braun2005direct,dreiser2008optical}, as well as theoretical predictions\cite{merkulov2002electron, khaetskii2002electron}. 

Comparing results from multiple QDs on the same device we find the amplitudes of electric field fluctuations to vary by about a factor two. Recent experimental works on the optical signatures of electric field fluctuations employing other techniques have shown a very wide range of both amplitudes and timescales, ranging from nanoseconds to seconds\cite{houel2012probing, nguyen2013photoneutralization, davancco2014multiple,matthiesen2014full} depending on the study. It appears the presence of charge noise depends sensitively upon material growth and device fabrication conditions rather than being an inherent property of QDs (in contrast to nuclear spin noise). In the present device the timescales of distinct electric field noise processes agree very well between QDs even if amplitudes vary; see Appendix B for additional data on QD B. The consistency of the timescales between QDs suggests the electric field fluctuations are due to distinct classes of charge traps present throughout the sample: the electric field noise dynamics, characterized by their the correlation times, are a global sample property. The noise amplitude for a particular QD, however, is a local property, which depends on the specific relative geometry of QD and noise sources.

\section{\label{sec:VI}Summary and conclusions}
In summary, we have investigated the contributions of nuclear spin bath fluctuations and dynamic electric field sources to the environmental noise of a QD in the presence of optical excitation. RFFS provides powerful tools to quantify these processes, for instance through bunching amplitudes of the intensity autocorrelation. Two-color excitation allows a clear distinction of the noise origins and permits unambiguous identification of nuclear bath correlation times. Two distinct correlation times associated with nuclear spin fluctuations are interpreted as arising from a partially shielded Knight field and hyperfine-mediated nuclear spin interaction. A separation of nuclear (\textless 50 $\mu$s) and electric field noise (\textgreater 1 ms) timescales makes a comparison to a Markovian model of time-averaged noise possible and allows us to quantify the environmental fluctuations. In the present sample, the dominant noise due to electric fields is described by a Gaussian distribution leading to spectral diffusion of 100-300 MHz while the Overhauser field magnitude corresponds to 22-25 mT at low excitation power. Our approach permits the direct quantitative comparison of individual QDs and different samples.

RFFS allows access to the rich physics of the central spin problem in the context of a confined system that is highly sensitive to both the inherent strain and interaction with a nearby Fermi sea. Exploring this parameter space in greater detail is the focus of future investigations. In addition, dynamics of the nuclear spin bath may be studied in the absence of an interacting electron\cite{maletinsky2007dynamics,chekhovich2014quadrupolar}. Here, the two-color excitation scheme which allows exclusive access to magnetic field fluctuations can be extended to neutral QDs, where the two transitions split by the fine structure are driven simultaneously. An extension of this work in a different direction could be studying the influence of feedback on the environment.

\section*{\label{ack}Acknowledgments}
We thank H. Ribeiro, R. Stockill, E. Chekhovich, A. Kuhlmann, R. Warburton and G. Solomon for useful discussions. C.M. gratefully acknowledges Clare College, Cambridge for financial support through a Junior Research Fellowship. We gratefully acknowledge financial support by the University of Cambridge, the European Research Council ERC Consolidator Grant agreement no. 617985 and the EU-FP7 Marie Curie Initial Training Network S$^{3}$NANO.

\appendix
\begin{figure}[tb]
\includegraphics[width=\columnwidth]{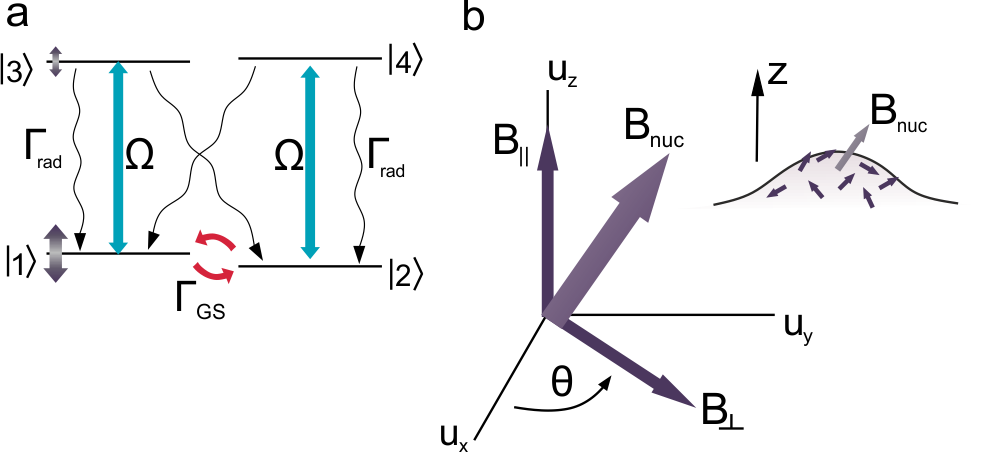}
\caption{\label{fig:6}(Color online) (a) Level structure and transitions for a negatively charged QD. Excited states decay radiatively, indicated by black wavy arrows, with rate $\Gamma_{\mathrm{rad}}$ and the ground state spin relaxes at $\Gamma_{\mathrm{GS}}$. In the absence of an external applied field the Zeeman splitting induced by the Overhauser field introduces a sub-linewidth splitting and consequently modifies the selection rules. (b) The instantaneous Overhauser field is decomposed in to cylindrical components in our model.}
\end{figure}
\section{Model of bunching amplitudes}
\subsection{\label{app:A1}X$^{-1}$ absorption in the presence of Overhauser field}
The intensity of resonance fluorescence is directly proportional to the excited state population, so we employ the optical Bloch equations to calculate this in the case of a negatively charged QD (X$^{1-}$ transition). The energy levels are indicated in Fig.~\ref{fig:6}. In the absence of an external applied magnetic field the degeneracy of the spin states is lifted due to the interaction with the ensemble of 10$^{4}$-10$^{5}$ nuclear spins in the QD. The hyperfine interaction is composed of a direct dipolar interactions between nuclear and electron/hole spins, and the dominant Fermi contact interaction term\cite{abragam1998principles}. For the heavy-hole wavefunctions in a QD, which are derived from underlying p-type orbitals, the interaction with the nuclear spins is of dipolar form and an order of magnitude smaller than the Fermi contact interaction with the electron spin\cite{fallahi2010measurement, chekhovich2011direct}; it is thus neglected. The Fermi contact hyperfine interaction is treated as an effective magnetic field (the Overhauser field) which provides the electron spin ground state quantization axis. In Faraday geometry where an external field is aligned with the growth axis, the ground state electron spin is quantized along this axis, where we represent these states of m$_{s}$ = $\pm$1/2 as $\Ket{\uparrow}$ = $\Ket{1}$ and $\Ket{\downarrow}$ = $\Ket{2}$. In this situation diagonal transitions are forbidden.
\begin{figure}[tb]
\includegraphics[width=\columnwidth]{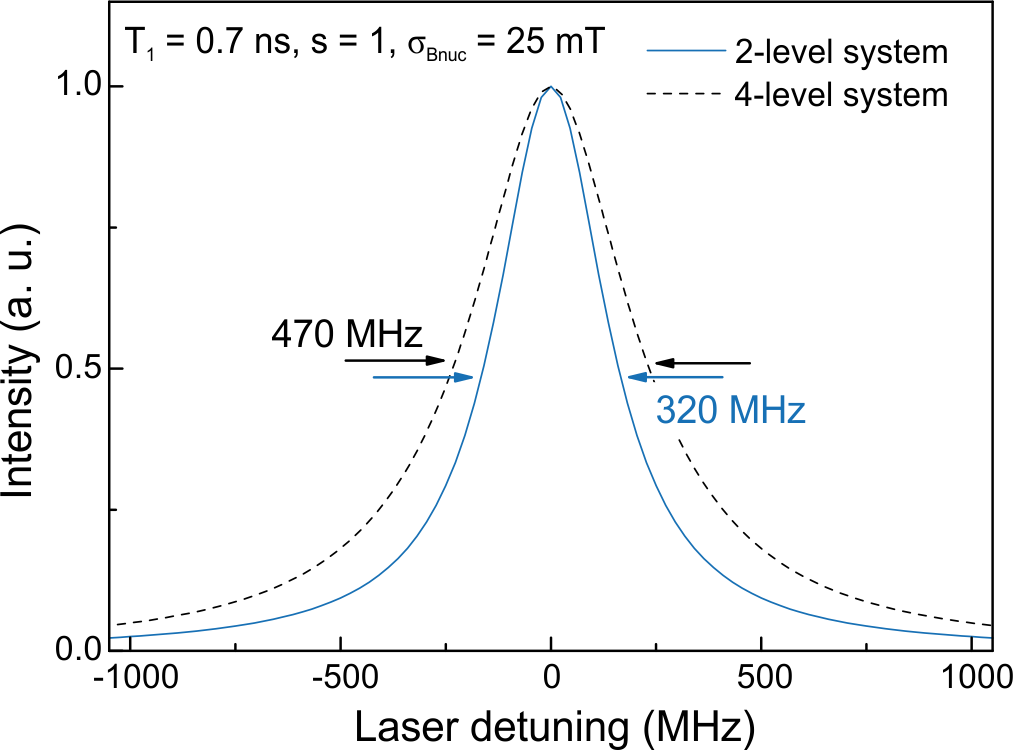}
\caption{\label{fig:7}(Color online) Absorption lineshape of an ideal two-level system (blue) and the X$^{1-}$ 4-level system with Gaussian Overhauser field of 25 mT standard deviation (black). The ground (4-level system only) and excited state lifetimes correspond to typical measured values.}
\end{figure}
Due to changes in the Overhauser field vector the electron ground state spin quantization axis shifts over time and hence the selection rules are not fixed. In general, the instantaneous eigenstates will be superpositions of the spin states $\Ket{\uparrow}$ and  $\Ket{\downarrow}$, allowing diagonal transitions for most Overhauser field configurations. The term of the Hamiltonian that describes the ground state coupling to the Overhauser field $\boldsymbol{B_{N}}$ can be expressed as
\begin{equation}
\begin{split}
\hat{\text{H}}_{\text{HF}} &= \frac{1}{2}\mu_{B}\text{g}_{e}\overrightarrow{\boldsymbol{B}}_{\boldsymbol{N}}\cdot\overrightarrow{\sigma}\\
&=\frac{1}{2}\mu_{B}\text{g}_{e}\Big( B_{\parallel}(\sigma_{11} - \sigma_{22}) + B_{\perp}(e^{i\theta}\sigma_{21} + e^{-i\theta}\sigma_{12})\Big)
\end{split}
\end{equation}
where we define the projection operators $\sigma_{ij}$ = $\Ket{i}\Bra{j}$, with i, j = 1$\dots$4 corresponding to one of  the four levels of X$^{1-}$. The angle $\theta$ is indicated in Fig. ~\ref{fig:6}.
The additional relevant physical parameters in our model are:
\begin{enumerate}
\item 	The spontaneous emission decay rate $\Gamma_{\mathrm{rad}}$ (see lifetime measurements in appendix B).
\item 	The QD-excitation field coupling strength given by the Rabi frequency. For convenience we use the parameter  $s = 2(\mathit{\Omega}/\Gamma_{\mathrm{rad}})^{2}$.
\item The ground state spin relaxation rate $\Gamma_{\mathrm{GS}}$.
\item 	The laser detuning from the transition frequency, $\delta$ = $\omega_{\mathrm{QD}} - \omega_{\mathrm{laser}}$.
\end{enumerate}
In general, pure dephasing (decay of coherences for reasons other than population decay) must also be considered. However, previous experiments on this sample have demonstrated slow pure dephasing rates\cite{matthiesen2012subnatural}, where measurements of the excited state coherence time suggested T$_{2}$ $\sim$ 2T$_{1}$ and so it is neglected in the discussion of the model that follows. We take into account both the electric dipole term representing interaction with the laser and the hyperfine term to write the Hamiltonian in the frame rotating at $\omega_{\mathrm{laser}}$ with respect to the laboratory frame:
\begin{equation}
\hat{\text{H}}_{\text{system}} = \hat{\text{H}}_{\text{HF}} + \hat{\text{H}}_{\text{dipole}}.
\end{equation}
The dipole interaction term can be written using the projection operators as:
\begin{equation}
 \hat{\text{H}}_{\text{dipole}} = \frac{1}{2}\hbar\mathit{\Omega}\Big (e^{-i\delta t}(\sigma_{13} + \sigma_{24}) + e^{i\delta t}(\sigma_{31} + \sigma_{42}) \Big).
\end{equation}
The time dependence of the resulting density matrix follows the Liouville von Neumann equation, 
\begin{equation}
\frac{d\rho}{dt} = -\frac{i}{\hbar}[\text{H}, \rho] + \mathlarger{\sum_{m}} \mathcal{L}(\rho, L_{m})
\end{equation}
where m = {1,2,3,4} and 
\begin{equation}
 \mathcal{L}(\rho, L_{m}) = L_{m}\rho L_{m}^{\dagger} - \frac{1}{2}\{ L_{m}^{\dagger}L_{m}, \rho\}.
\end{equation}
Relaxation of the ground state spin ($\Gamma_{\textrm{GS}}$) and spontaneous emission processes ($\Gamma_{\textrm{rad}}$) are included in the Lindblad operators:
\begin{subequations}
\begin{eqnarray}
L_{1}& =&(\Gamma_{\mathrm{GS}})^{\frac{1}{2}} \sigma_{12}, \\
L_{2}& =&(\Gamma_{\mathrm{GS}})^{\frac{1}{2}} \sigma_{21},\\
L_{3}& =&(\Gamma_{\mathrm{rad}})^{\frac{1}{2}} \sigma_{13},\\
L_{4}& =&(\Gamma_{\mathrm{rad}})^{\frac{1}{2}} \sigma_{24}.
\end{eqnarray}
\end{subequations}
We measure resonance fluorescence intensity on timescales longer than the radiative lifetime. Therefore we are interested in the excited state population $\rho_{33}+ \rho_{44}$ in the stationary limit $\frac{d\rho}{dt} = 0$. 

Next, we illustrate the effect of hyperfine coupling on the optical properties of the QD by considering the resulting absorption lineshape. Figure~\ref{fig:7} compares the absorption lineshapes expected for an ideal two-level system (blue curve) and the four-level QD system (black curve) given an Overhauser field distribution with finite variance. Typical values are chosen for the parameters discussed above:
\begin{enumerate}
\item Spontaneous emission rate $\Gamma_{\mathrm{rad}}$  = $(2\pi \mathrm{T_{1}})^{-1}$, T$_{1}$ = 700 ps.
\item Saturation parameter $s = 1$.
\item Ground state relaxation rate $\Gamma_{\mathrm{GS}}$ = 2 x 10$^{-4}$ s$^{-1}$.
\item Overhauser field standard deviation $\sigma_{\mathrm{B}}$ = 25 mT.
\end{enumerate}
For the two-level system the curve represents the expected Lorentzian power-broadened lineshape. Interestingly, in the case of the X$^{1-}$ level structure, we obtain a Lorentzian lineshape again, albeit broadened. The amplitude of both curves, corresponding to the intensity of resonance fluorescence, has been scaled to unity here, while in actual fact, the intensity is reduced in the four-level case, mainly due to spin pumping.

\subsection{\label{app:A2}Autocorrelation bunching amplitudes}
The intensity autocorrelation of a time-binned signal written as $\{x_{1}, x_{2}, \dots, x_{\mathrm{N}} \}$  with mean $\langle I(t)\rangle = \bar{x}$ has a zero time delay amplitude given by: 
\begin{equation}
g^{(2)}(0) = \frac{1}{N}\frac{\sum_{i}x_{i}^{2}}{\bar{x}^{2}}.
\end{equation}
This can be written directly in terms of the variance $\sigma^{2}$ and the mean as
\begin{equation}
g^{(2)}(0) - 1 = \frac{\sigma^{2}}{\bar{x}^{2}}.
\end{equation}
We therefore may relate the variance of our entire signal time trace to the full amplitude of the autocorrelation. 

In the following we will be considering the autocorrelation amplitude of electric field noise in particular. Given the clear division of nuclear spin and electric field-related timescales found experimentally we can consider a cut-off time in the autocorrelation that separates the two. The autocorrelation amplitude at this cut-off point then captures all fluctuations due to electric field noise.
\subsection{\label{app:A3}Model of electric and Overhauser field distributions}
We assume that during the measurement time the full range of possible electric field values is explored. The transition frequency distribution $P(\Delta\delta)$ is represented by a Gaussian distribution about a central resonant frequency:
\begin{equation}
\label{eq:PE}
P(\Delta\delta) = \frac{1}{\sqrt{2\pi \sigma_{E}^{2}}}\mathrm{exp}\bigg[ - \frac{1}{2}\Big( \frac{\Delta\delta}{\sigma_{E}}\Big)^{2}\bigg].
\end{equation}
Here $\Delta\delta$ is the detuning with respect to the central frequency arising from the electric-field induced Stark shifts. The distribution has a standard deviation, $\sigma_{E}$, corresponding to a full-width at half-maximum $\Delta_{\mathrm{FWHM}} = \sqrt{8\ln 2} \sigma_{E}$. 

For the Overhauser field vector we choose an isotropic Gaussian distribution
\begin{equation}
W(\boldsymbol{B_{N}}) = \frac{1}{(2\pi\sigma_{B}^{2})^{3/2}}\mathrm{exp}\bigg[ - \frac{1}{2}\Big( \frac{\boldsymbol{B_{N}}}{\sigma_{B}} \Big)^{2}  \bigg],
\end{equation}
where $\boldsymbol{B_{N}}$ is the instantaneous Overhauser field vector and $\sigma_{B}$ is the standard deviation of the field \cite{merkulov2002electron,urbaszek2013nuclear}. When considering the combined effects of the two noise sources we take advantage of the separation of timescales and calculate the time-averaged effect of a fluctuating Overhauser field, that is, an absorption lineshape such as found in Fig. ~\ref{fig:7}. Resonance fluorescence noise amplitudes due to electric field fluctuations are then obtained by allowing the central frequency of the QD transition to vary according to the probability distribution in Eq. \eqref{eq:PE}.

\subsection{\label{app:A4}Parameters relevant to the model}
The sensitivity to electric and Overhauser field fluctuations is determined by the underlying Lorentzian absorption spectrum of a QD transition. The radiative lifetime $\text{T}_{1}$ gives directly the natural linewidth of the transition, where power broadening produces the linewidth under excitation, $\Gamma_{\mathrm{FWHM}} = (2\pi \text{T}_{1})^{-1}\sqrt{1+s}$. Consequently, the sensitivity to both electric and magnetic field noise drops rapidly as the QD transition is saturated. To model the bunching amplitudes for a particular QD it is necessary to measure both the radiative lifetime and saturation behavior of every QD.

\subsection{\label{app:A5}Additional data fit to the electric field amplitude model}
\begin{figure}[tb]
\includegraphics[width=\columnwidth]{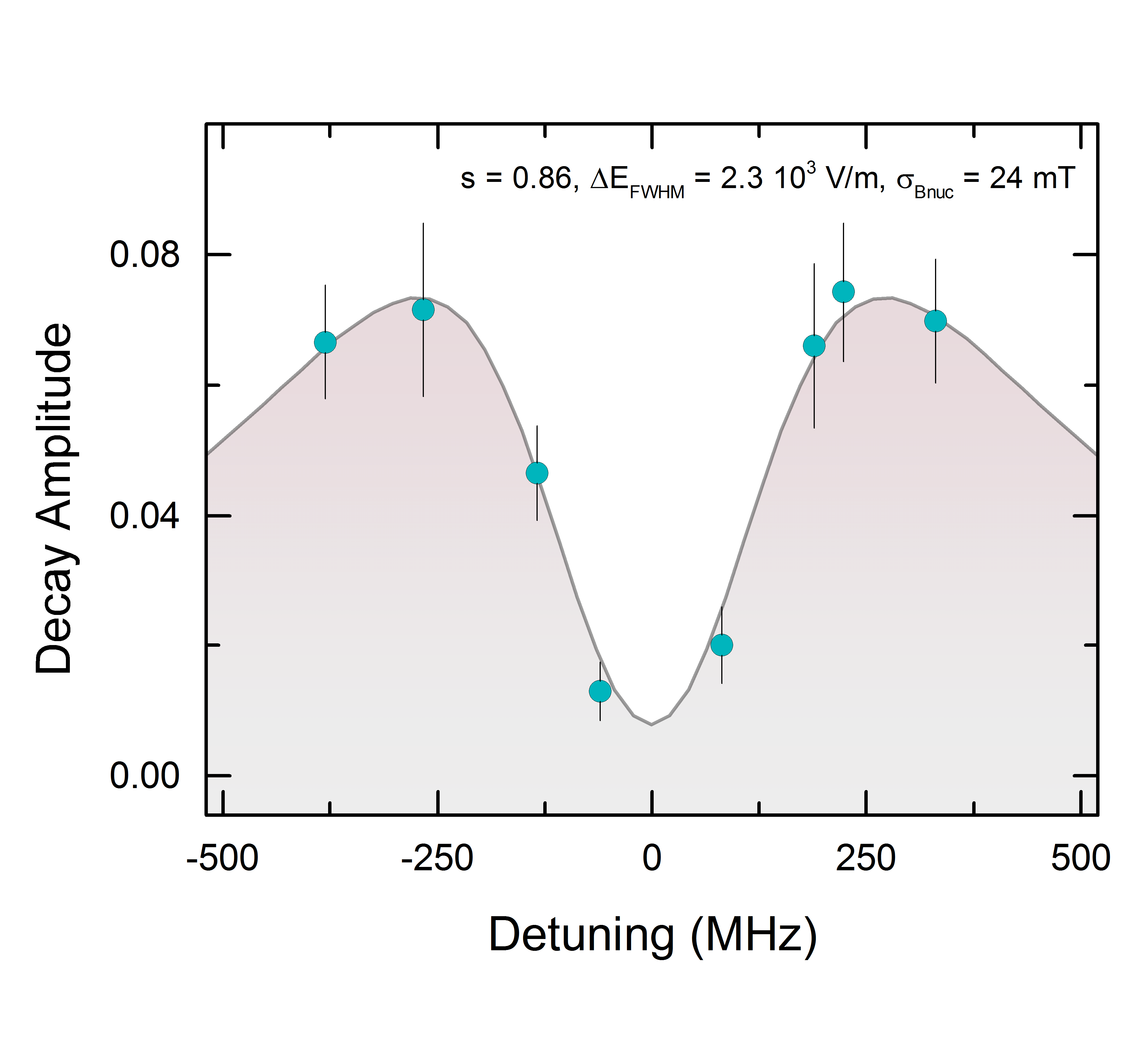}
\caption{\label{fig:8}(Color online) Fit of electric field noise amplitudes to the model for QD C, where the QD is driven with saturation parameter $s=0.86$. The extracted standard deviation for the underlying Overhauser field is 24 mT, and the Gaussian distribution describing the shifts in resonant frequency due to the electric field has a full width at half maximum of (147$\pm$5) MHz.}
\end{figure}
Here we present an example of the model applied to QD C. The sum of electric field noise amplitudes and a fit is shown in Fig.~\ref{fig:8}. We extract an Overhauser field distribution with a standard deviation of $(24\pm2)$ mT. The error in the fit presented in Fig.~\ref{fig:8} is minimized for electric field noise with an $\Delta E_{\mathrm{FWHM}}$ of $(2.3\pm0.1)10^{3}$ V/m or, equivalently, a transition broadening of $(147\pm5)$ MHz.

\begin{figure}[bht]
\includegraphics[width=\columnwidth]{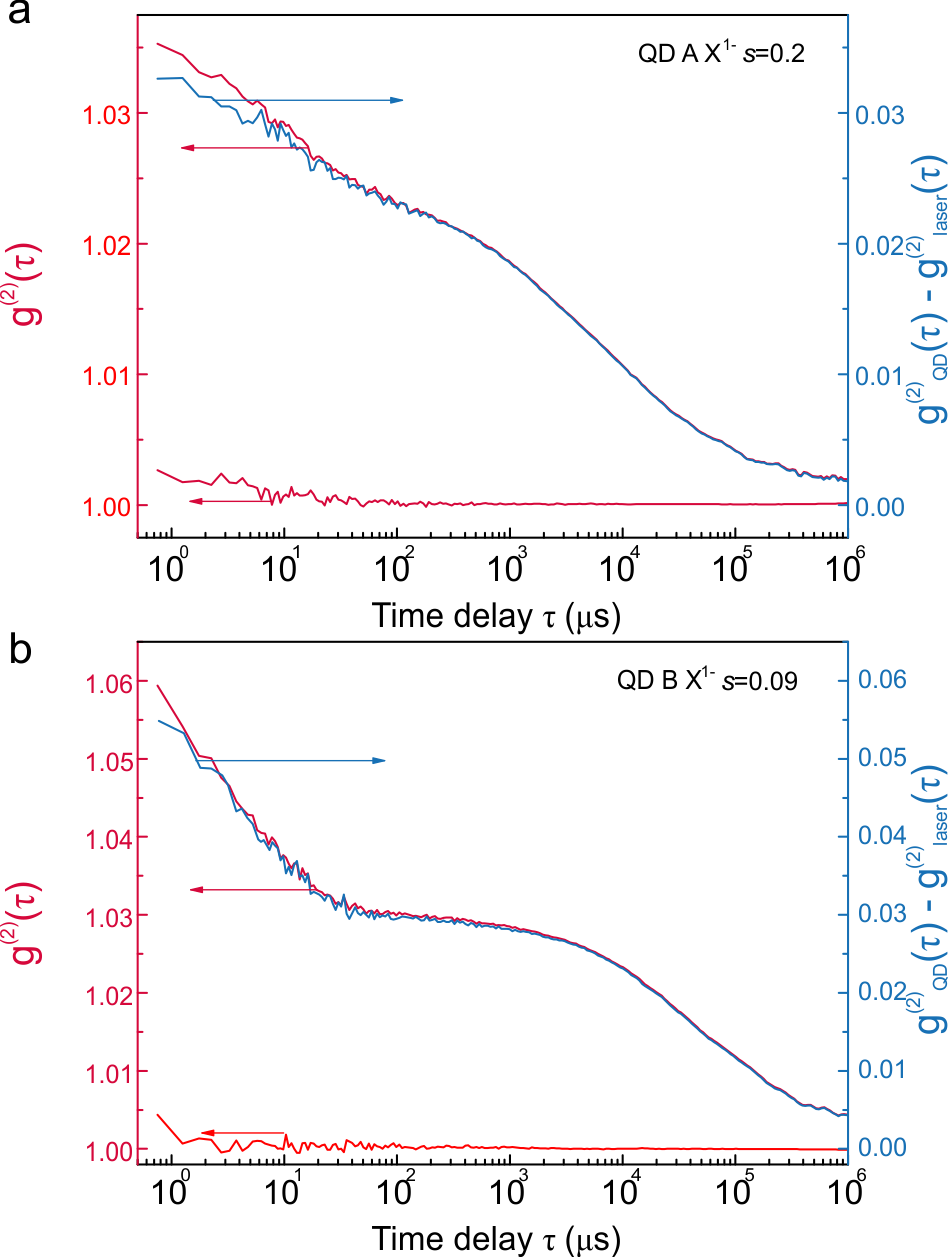}
\caption{\label{fig:9}(Color online) Data treatment for calculated autocorrelation functions. Red curves show the raw autocorrelations for both QD RF and laser output at a comparable count rate. The blue curve is the difference between QD RF and laser autocorrelations, and accounts for intensity fluctuations inherent to the measurement apparatus, such as detector afterpulsing. (a) and (b) represent two examples recorded for different parameters of the measurement system such as detector count rate, laser frequency and power incident upon the experimental set-up.}
\end{figure}

\subsection{\label{app:A6}Fit to autocorrelation bunching decays}
Data is fit to a sum of multiple exponential decays in order to extract rates of noise processes. Extracting individual amplitudes is useful to identify the origins of each component of noise.  In appendix B we present further correlation timescales and amplitudes found for QD B. 

A single exponential decay with correlation time $\tau_{c}$ is indicative of a single relaxation process, where the corresponding power spectrum (directly related via the Wiener-Khinchin theorem) is a single Lorentzian peak with a width that is proportional to $1/\tau_{c}$\cite{machlup1954noise}. In our data we are able to consistently extract between 4 and 6 exponential decays, which suggests this is the number of distinct processes contributing to noise. We note that, in addition to exponential decays in the autocorrelations, a 1/f component is also present at low frequencies.
We model electric field fluctuations by a Gaussian distribution of resonant frequencies which is a good description for the effect of noise upon photon counting statistics\cite{matthiesen2014full}. However, a Gaussian distribution is consistent with a large number of electric field values, not initially in keeping with a small number of charge traps. One picture is that a relatively small number of independently fluctuating charge traps, N, which can be occupied or unoccupied, leads to $2^{\mathrm{N}}$ possible electric field values at the position of the dot. In addition, single decay timescales in the autocorrelation may be associated with many similar charge traps rather than single locations, potentially increasing N. There is also the possibility that the charge traps interact; in this case a large number of traps with a range of associated timescales again result in Lorentzian noise spectra and thus exponential decays in autocorrelations\cite{hooge1997correlation}. 

\section{Supporting data}
\subsection{\label{app:B1}Background correction of data}
Figure~\ref{fig:9} shows the treatment of measured autocorrelation data for two examples. The autocorrelation function for the detection of laser emission at comparable count rate is subtracted from the autocorrelation measured for QD fluorescence. This background data is taken with the QD transition detuned from the resonant laser, where the polarization suppression is relaxed to gain the same photon count rates. While APD afterpulsing has a pronounced effect at time delays up to about 1 $\mu\mathrm{s}$, small corrections resulting from the subtractions are visible for time delays as large as 100 $\mu\mathrm{s}$, rendering it necessary to take into account background for all data. Further, the autocorrelation function depends sensitively on experimental settings, such as APD count rate or laser power stabilization, and changes when equipment is exchanged. For this reason the reference measurement of the laser autocorrelation has to replicate experimental conditions as closely as possible.

\begin{figure}[tb]
\includegraphics[width=\columnwidth]{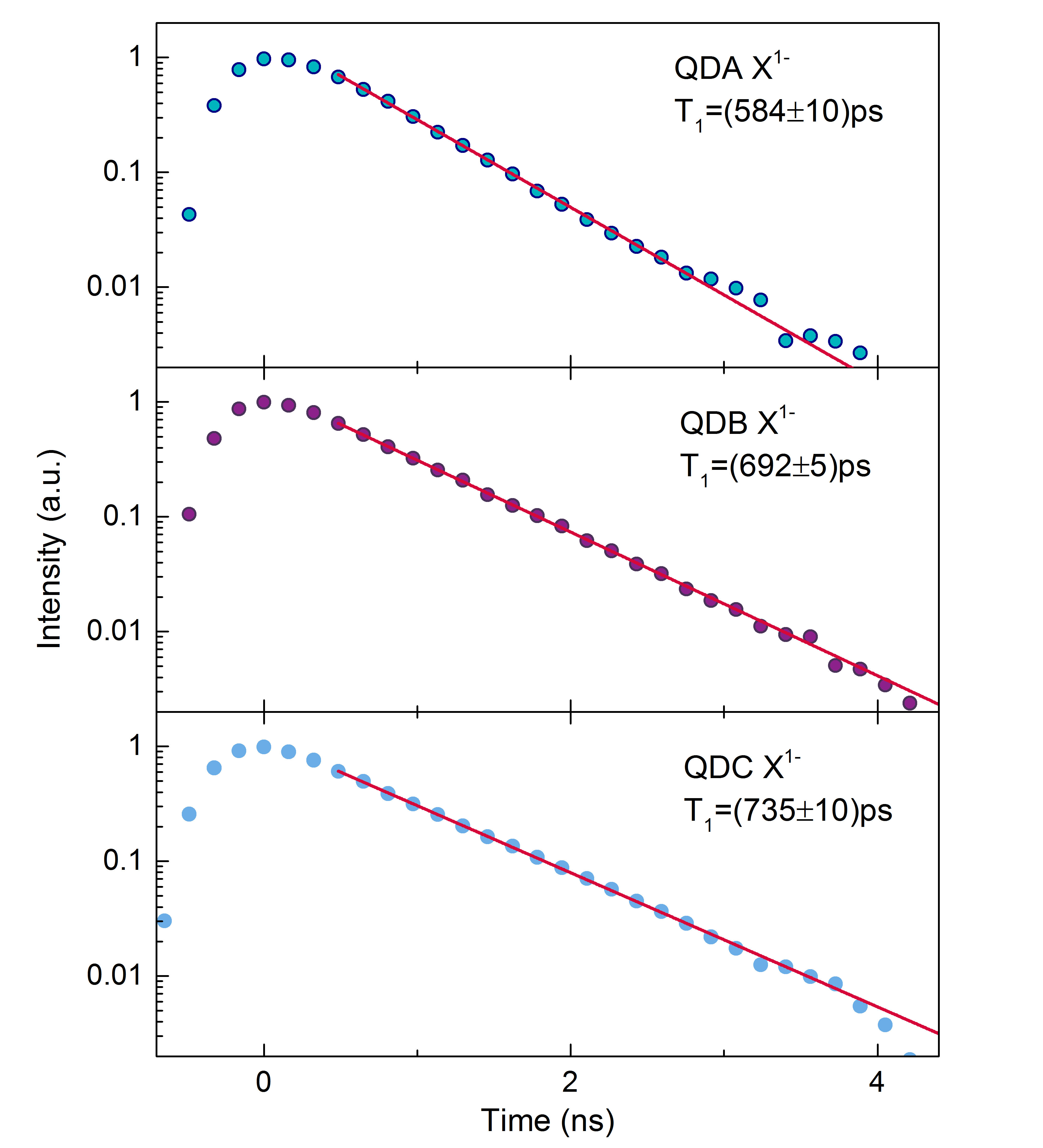}
\caption{\label{fig:10}(Color online) Measurements of the radiative lifetime for QDs A, B, C using time-correlated single photon counting under pulsed resonant excitation.}
\end{figure}

\subsection{\label{app:B2}Lifetime measurements of QD A, B, C}
The excited state lifetime $\text{T}_{1}$ is measured under pulsed resonant excitation, using an electro-optic modulator with 10 GHz bandwidth driven by voltage pulses with sub-50 ps rise and fall times. QD resonance fluorescence detection times are recorded in bins of 162 ps width with respect to a trigger signal derived from the pulsed voltage source. Data for the three QDs used in this paper are plotted in Fig.~\ref{fig:10}, together with single exponential fit functions. The error is the standard error in the mean for independent fits to decay curves under repeated measurement (four for QDs A and C and eight for QD B). The timing resolution of the measurement system amounts to $\sim$ 350 ps.

\subsection{\label{app:B3}Detailed autocorrelation amplitudes and timescales for QD B}
\begin{figure}[tb]
\includegraphics[width=\columnwidth]{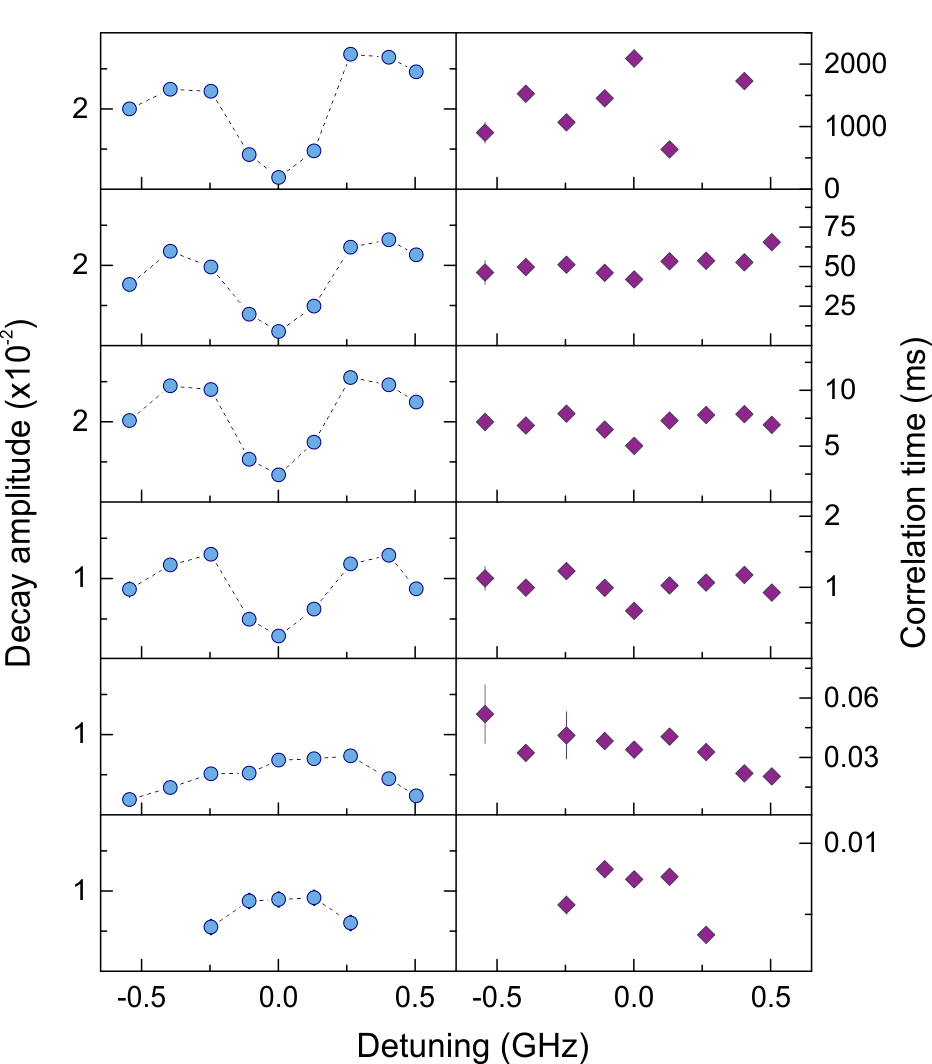}
\caption{\label{fig:11}(Color online) Autocorrelation decay amplitudes and timescales for QD B, $s$ = 0.92. The lower two panels are identified as noise due to nuclear spin fluctuations, whilst the upper four panels show a detuning dependence consistent with underlying electric field fluctuations, as discussed in the main text. All nuclear spin fluctuation timescales are again below 100 $\mu$s, whilst electric field fluctuations persist from 1 ms up to seconds.}
\end{figure}
Figure~\ref{fig:11} displays amplitudes and timescales as extracted from exponential fits to the measured autocorrelation functions of QD B. As described in the main text the bottom two sets of panels represent nuclear field fluctuations while the top panels describe resonance fluorescence fluctuations due to electric field noise. We calculate correlation times up to $\sim$ 1 s from the resonance fluorescence time traces as before, but note that noise processes (due to electric field fluctuations) with considerably longer correlation times take place in our samples as well. These slow dynamics can be accessed in measurements with long acquisition times, but are unlikely to differ qualitatively from the electric-field noise we observe on faster timescales.

In comparison to data for QD A, cf. Fig.~\ref{fig:2}, decay amplitudes related to electric field fluctuations are reduced by about a factor for QD B. However, we note the timescales are very similar in the two cases and are consistent with the values measured for other QDs of the same sample, as expected when the noise arises from sample-dependent defects.

\subsection{\label{app:B4}Sample structure}
\begin{figure*}[tb]
\includegraphics[width=1.5\columnwidth]{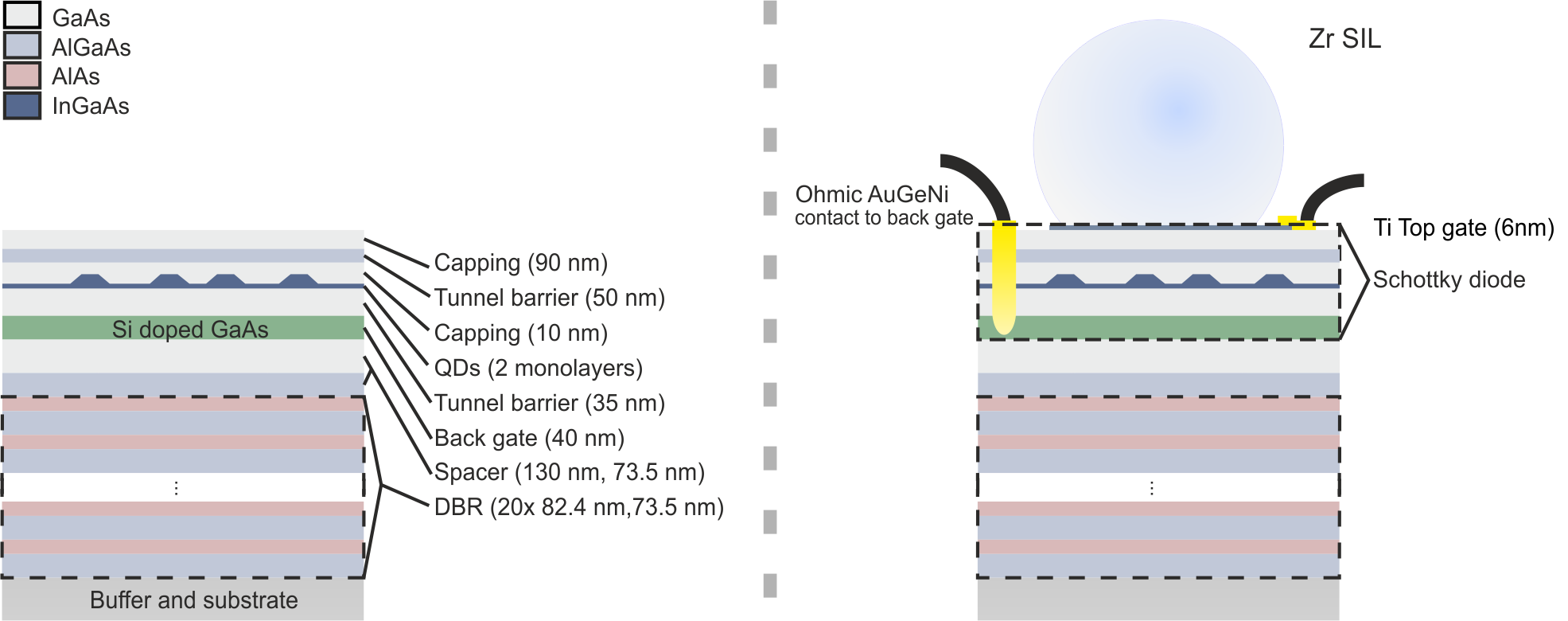}
\caption{\label{fig:12}(Color online) Left: the sample structure, indicating all MBE-grown layers. Right: post-growth Ohmic and Schottky contacts are applied to the diode structure and a SIL is placed on the sample surface.}
\end{figure*}
Our sample structure is illustrated in Fig.~\ref{fig:12}.  Self-assembled InGaAs QDs are incorporated into a Schottky diode structure with a 35-nm tunnel barrier between the QD layer and an n-doped layer. The diode heterostructure is grown above a distributed Bragg reflector to maximize photon outcoupling efficiency. Further enhancement of photon collection is obtained by the presence of a super-hemispherical solid immersion lens placed directly on the semi-transparent Titanium Schottky contact on the surface of the sample. For the current sample we estimate a photon outcoupling efficiency of up to 15 \% for QDs with emission wavelengths around 970-980 nm.
\bibliography{references}

\begin{thebibliography}{46}
\expandafter\ifx\csname natexlab\endcsname\relax\def\natexlab#1{#1}\fi
\expandafter\ifx\csname bibnamefont\endcsname\relax
  \def\bibnamefont#1{#1}\fi
\expandafter\ifx\csname bibfnamefont\endcsname\relax
  \def\bibfnamefont#1{#1}\fi
\expandafter\ifx\csname citenamefont\endcsname\relax
  \def\citenamefont#1{#1}\fi
\expandafter\ifx\csname url\endcsname\relax
  \def\url#1{\texttt{#1}}\fi
\expandafter\ifx\csname urlprefix\endcsname\relax\def\urlprefix{URL }\fi
\providecommand{\bibinfo}[2]{#2}
\providecommand{\eprint}[2][]{\url{#2}}

\bibitem[{\citenamefont{Gammon and Steel}(2002)}]{gammon2002optical}
\bibinfo{author}{\bibfnamefont{D.}~\bibnamefont{Gammon}} \bibnamefont{and}
  \bibinfo{author}{\bibfnamefont{D.~G.} \bibnamefont{Steel}},
  \bibinfo{journal}{Physics Today} \textbf{\bibinfo{volume}{55}},
  \bibinfo{pages}{36} (\bibinfo{year}{2002}).

\bibitem[{\citenamefont{Warburton et~al.}(2002)\citenamefont{Warburton,
  Schulhauser, Haft, Sch{\"a}flein, Karrai, Garcia, Schoenfeld, and
  Petroff}}]{warburton2002giant}
\bibinfo{author}{\bibfnamefont{R.~J.} \bibnamefont{Warburton}},
  \bibinfo{author}{\bibfnamefont{C.}~\bibnamefont{Schulhauser}},
  \bibinfo{author}{\bibfnamefont{D.}~\bibnamefont{Haft}},
  \bibinfo{author}{\bibfnamefont{C.}~\bibnamefont{Sch{\"a}flein}},
  \bibinfo{author}{\bibfnamefont{K.}~\bibnamefont{Karrai}},
  \bibinfo{author}{\bibfnamefont{J.~M.} \bibnamefont{Garcia}},
  \bibinfo{author}{\bibfnamefont{W.}~\bibnamefont{Schoenfeld}},
  \bibnamefont{and} \bibinfo{author}{\bibfnamefont{P.~M.}
  \bibnamefont{Petroff}}, \bibinfo{journal}{Physical Review B}
  \textbf{\bibinfo{volume}{65}}, \bibinfo{pages}{113303}
  (\bibinfo{year}{2002}).

\bibitem[{\citenamefont{Bayer et~al.}(2002)\citenamefont{Bayer, Ortner, Stern,
  Kuther, Gorbunov, Forchel, Hawrylak, Fafard, Hinzer, Reinecke
  et~al.}}]{bayer2002fine}
\bibinfo{author}{\bibfnamefont{M.}~\bibnamefont{Bayer}},
  \bibinfo{author}{\bibfnamefont{G.}~\bibnamefont{Ortner}},
  \bibinfo{author}{\bibfnamefont{O.}~\bibnamefont{Stern}},
  \bibinfo{author}{\bibfnamefont{A.}~\bibnamefont{Kuther}},
  \bibinfo{author}{\bibfnamefont{A.}~\bibnamefont{Gorbunov}},
  \bibinfo{author}{\bibfnamefont{A.}~\bibnamefont{Forchel}},
  \bibinfo{author}{\bibfnamefont{P.}~\bibnamefont{Hawrylak}},
  \bibinfo{author}{\bibfnamefont{S.}~\bibnamefont{Fafard}},
  \bibinfo{author}{\bibfnamefont{K.}~\bibnamefont{Hinzer}},
  \bibinfo{author}{\bibfnamefont{T.}~\bibnamefont{Reinecke}},
  \bibnamefont{et~al.}, \bibinfo{journal}{Physical Review B}
  \textbf{\bibinfo{volume}{65}}, \bibinfo{pages}{195315}
  (\bibinfo{year}{2002}).

\bibitem[{\citenamefont{Finley et~al.}(2002)\citenamefont{Finley, Mowbray,
  Skolnick, Ashmore, Baker, Monte, and Hopkinson}}]{finley2002fine}
\bibinfo{author}{\bibfnamefont{J.}~\bibnamefont{Finley}},
  \bibinfo{author}{\bibfnamefont{D.}~\bibnamefont{Mowbray}},
  \bibinfo{author}{\bibfnamefont{M.}~\bibnamefont{Skolnick}},
  \bibinfo{author}{\bibfnamefont{A.}~\bibnamefont{Ashmore}},
  \bibinfo{author}{\bibfnamefont{C.}~\bibnamefont{Baker}},
  \bibinfo{author}{\bibfnamefont{A.}~\bibnamefont{Monte}}, \bibnamefont{and}
  \bibinfo{author}{\bibfnamefont{M.}~\bibnamefont{Hopkinson}},
  \bibinfo{journal}{Physical Review B} \textbf{\bibinfo{volume}{66}},
  \bibinfo{pages}{153316} (\bibinfo{year}{2002}).

\bibitem[{\citenamefont{Vamivakas et~al.}(2011)\citenamefont{Vamivakas, Zhao,
  F{\"a}lt, Badolato, Taylor, and Atat{\"u}re}}]{vamivakas2011nanoscale}
\bibinfo{author}{\bibfnamefont{A.}~\bibnamefont{Vamivakas}},
  \bibinfo{author}{\bibfnamefont{Y.}~\bibnamefont{Zhao}},
  \bibinfo{author}{\bibfnamefont{S.}~\bibnamefont{F{\"a}lt}},
  \bibinfo{author}{\bibfnamefont{A.}~\bibnamefont{Badolato}},
  \bibinfo{author}{\bibfnamefont{J.}~\bibnamefont{Taylor}}, \bibnamefont{and}
  \bibinfo{author}{\bibfnamefont{M.}~\bibnamefont{Atat{\"u}re}},
  \bibinfo{journal}{Physical Review Letters} \textbf{\bibinfo{volume}{107}},
  \bibinfo{pages}{166802} (\bibinfo{year}{2011}).

\bibitem[{\citenamefont{Wilson-Rae et~al.}(2004)\citenamefont{Wilson-Rae,
  Zoller, and Imamoḡlu}}]{wilson2004laser}
\bibinfo{author}{\bibfnamefont{I.}~\bibnamefont{Wilson-Rae}},
  \bibinfo{author}{\bibfnamefont{P.}~\bibnamefont{Zoller}}, \bibnamefont{and}
  \bibinfo{author}{\bibfnamefont{A.}~\bibnamefont{Imamoḡlu}},
  \bibinfo{journal}{Physical Review Letters} \textbf{\bibinfo{volume}{92}},
  \bibinfo{pages}{075507} (\bibinfo{year}{2004}).

\bibitem[{\citenamefont{Yeo et~al.}(2014)\citenamefont{Yeo, de~Assis, Gloppe,
  Dupont-Ferrier, Verlot, Malik, Dupuy, Claudon, G{\'e}rard, Auff{\`e}ves
  et~al.}}]{yeo2013strain}
\bibinfo{author}{\bibfnamefont{I.}~\bibnamefont{Yeo}},
  \bibinfo{author}{\bibfnamefont{P.-L.} \bibnamefont{de~Assis}},
  \bibinfo{author}{\bibfnamefont{A.}~\bibnamefont{Gloppe}},
  \bibinfo{author}{\bibfnamefont{E.}~\bibnamefont{Dupont-Ferrier}},
  \bibinfo{author}{\bibfnamefont{P.}~\bibnamefont{Verlot}},
  \bibinfo{author}{\bibfnamefont{N.~S.} \bibnamefont{Malik}},
  \bibinfo{author}{\bibfnamefont{E.}~\bibnamefont{Dupuy}},
  \bibinfo{author}{\bibfnamefont{J.}~\bibnamefont{Claudon}},
  \bibinfo{author}{\bibfnamefont{J.-M.} \bibnamefont{G{\'e}rard}},
  \bibinfo{author}{\bibfnamefont{A.}~\bibnamefont{Auff{\`e}ves}},
  \bibnamefont{et~al.}, \bibinfo{journal}{Nature Nanotechnology}
  \textbf{\bibinfo{volume}{9}}, \bibinfo{pages}{106} (\bibinfo{year}{2014}).

\bibitem[{\citenamefont{Montinaro et~al.}(2014)\citenamefont{Montinaro,
  W{\"u}st, Munsch, Fontana, Russo-Averchi, Heiss, Fontcuberta~i Morral,
  Warburton, and Poggio}}]{montinaro2014quantum}
\bibinfo{author}{\bibfnamefont{M.}~\bibnamefont{Montinaro}},
  \bibinfo{author}{\bibfnamefont{G.}~\bibnamefont{W{\"u}st}},
  \bibinfo{author}{\bibfnamefont{M.}~\bibnamefont{Munsch}},
  \bibinfo{author}{\bibfnamefont{Y.}~\bibnamefont{Fontana}},
  \bibinfo{author}{\bibfnamefont{E.}~\bibnamefont{Russo-Averchi}},
  \bibinfo{author}{\bibfnamefont{M.}~\bibnamefont{Heiss}},
  \bibinfo{author}{\bibfnamefont{A.}~\bibnamefont{Fontcuberta~i Morral}},
  \bibinfo{author}{\bibfnamefont{R.~J.} \bibnamefont{Warburton}},
  \bibnamefont{and} \bibinfo{author}{\bibfnamefont{M.}~\bibnamefont{Poggio}},
  \bibinfo{journal}{arXiv preprint arXiv:1405.2821v1}  (\bibinfo{year}{2014}).

\bibitem[{\citenamefont{Warburton}(2013)}]{warburton2013single}
\bibinfo{author}{\bibfnamefont{R.~J.} \bibnamefont{Warburton}},
  \bibinfo{journal}{Nature Materials} \textbf{\bibinfo{volume}{12}},
  \bibinfo{pages}{483} (\bibinfo{year}{2013}).

\bibitem[{\citenamefont{Lodahl et~al.}(2013)\citenamefont{Lodahl, Mahmoodian,
  and Stobbe}}]{lodahl2013interfacing}
\bibinfo{author}{\bibfnamefont{P.}~\bibnamefont{Lodahl}},
  \bibinfo{author}{\bibfnamefont{S.}~\bibnamefont{Mahmoodian}},
  \bibnamefont{and} \bibinfo{author}{\bibfnamefont{S.}~\bibnamefont{Stobbe}},
  \bibinfo{journal}{arXiv preprint arXiv:1312.1079}  (\bibinfo{year}{2013}).

\bibitem[{\citenamefont{Abragam}(1961)}]{abragam1998principles}
\bibinfo{author}{\bibfnamefont{A.}~\bibnamefont{Abragam}},
  \emph{\bibinfo{title}{Principles of nuclear magnetism}}
  (\bibinfo{publisher}{Oxford University Press}, \bibinfo{address}{Oxford},
  \bibinfo{year}{1961}).

\bibitem[{\citenamefont{Urbaszek et~al.}(2013)\citenamefont{Urbaszek, Marie,
  Amand, Krebs, Voisin, Maletinsky, H{\"o}gele, and
  Imamoglu}}]{urbaszek2013nuclear}
\bibinfo{author}{\bibfnamefont{B.}~\bibnamefont{Urbaszek}},
  \bibinfo{author}{\bibfnamefont{X.}~\bibnamefont{Marie}},
  \bibinfo{author}{\bibfnamefont{T.}~\bibnamefont{Amand}},
  \bibinfo{author}{\bibfnamefont{O.}~\bibnamefont{Krebs}},
  \bibinfo{author}{\bibfnamefont{P.}~\bibnamefont{Voisin}},
  \bibinfo{author}{\bibfnamefont{P.}~\bibnamefont{Maletinsky}},
  \bibinfo{author}{\bibfnamefont{A.}~\bibnamefont{H{\"o}gele}},
  \bibnamefont{and} \bibinfo{author}{\bibfnamefont{A.}~\bibnamefont{Imamoglu}},
  \bibinfo{journal}{Reviews of Modern Physics} \textbf{\bibinfo{volume}{85}},
  \bibinfo{pages}{79} (\bibinfo{year}{2013}).

\bibitem[{\citenamefont{Chekhovich et~al.}(2013)\citenamefont{Chekhovich,
  Makhonin, Tartakovskii, Yacoby, Bluhm, Nowack, and
  Vandersypen}}]{chekhovich2013nuclear}
\bibinfo{author}{\bibfnamefont{E.}~\bibnamefont{Chekhovich}},
  \bibinfo{author}{\bibfnamefont{M.}~\bibnamefont{Makhonin}},
  \bibinfo{author}{\bibfnamefont{A.}~\bibnamefont{Tartakovskii}},
  \bibinfo{author}{\bibfnamefont{A.}~\bibnamefont{Yacoby}},
  \bibinfo{author}{\bibfnamefont{H.}~\bibnamefont{Bluhm}},
  \bibinfo{author}{\bibfnamefont{K.}~\bibnamefont{Nowack}}, \bibnamefont{and}
  \bibinfo{author}{\bibfnamefont{L.}~\bibnamefont{Vandersypen}},
  \bibinfo{journal}{Nature Materials} \textbf{\bibinfo{volume}{12}},
  \bibinfo{pages}{494} (\bibinfo{year}{2013}).

\bibitem[{\citenamefont{Merkulov et~al.}(2002)\citenamefont{Merkulov, Efros,
  and Rosen}}]{merkulov2002electron}
\bibinfo{author}{\bibfnamefont{I.}~\bibnamefont{Merkulov}},
  \bibinfo{author}{\bibfnamefont{A.~L.} \bibnamefont{Efros}}, \bibnamefont{and}
  \bibinfo{author}{\bibfnamefont{M.}~\bibnamefont{Rosen}},
  \bibinfo{journal}{Physical Review B} \textbf{\bibinfo{volume}{65}},
  \bibinfo{pages}{205309} (\bibinfo{year}{2002}).

\bibitem[{\citenamefont{Khaetskii et~al.}(2002)\citenamefont{Khaetskii, Loss,
  and Glazman}}]{khaetskii2002electron}
\bibinfo{author}{\bibfnamefont{A.~V.} \bibnamefont{Khaetskii}},
  \bibinfo{author}{\bibfnamefont{D.}~\bibnamefont{Loss}}, \bibnamefont{and}
  \bibinfo{author}{\bibfnamefont{L.}~\bibnamefont{Glazman}},
  \bibinfo{journal}{Physical Review Letters} \textbf{\bibinfo{volume}{88}},
  \bibinfo{pages}{186802} (\bibinfo{year}{2002}).

\bibitem[{\citenamefont{Coish and Loss}(2004)}]{coish2004hyperfine}
\bibinfo{author}{\bibfnamefont{W.}~\bibnamefont{Coish}} \bibnamefont{and}
  \bibinfo{author}{\bibfnamefont{D.}~\bibnamefont{Loss}},
  \bibinfo{journal}{Physical Review B} \textbf{\bibinfo{volume}{70}},
  \bibinfo{pages}{195340} (\bibinfo{year}{2004}).

\bibitem[{\citenamefont{Cywi{\'n}ski et~al.}(2009)\citenamefont{Cywi{\'n}ski,
  Witzel, and Sarma}}]{cywinski2009electron}
\bibinfo{author}{\bibfnamefont{{\L}.}~\bibnamefont{Cywi{\'n}ski}},
  \bibinfo{author}{\bibfnamefont{W.~M.} \bibnamefont{Witzel}},
  \bibnamefont{and} \bibinfo{author}{\bibfnamefont{S.~D.} \bibnamefont{Sarma}},
  \bibinfo{journal}{Physical Review Letters} \textbf{\bibinfo{volume}{102}},
  \bibinfo{pages}{057601} (\bibinfo{year}{2009}).

\bibitem[{\citenamefont{Sinitsyn et~al.}(2012)\citenamefont{Sinitsyn, Li,
  Crooker, Saxena, and Smith}}]{sinitsyn2012role}
\bibinfo{author}{\bibfnamefont{N.}~\bibnamefont{Sinitsyn}},
  \bibinfo{author}{\bibfnamefont{Y.}~\bibnamefont{Li}},
  \bibinfo{author}{\bibfnamefont{S.}~\bibnamefont{Crooker}},
  \bibinfo{author}{\bibfnamefont{A.}~\bibnamefont{Saxena}}, \bibnamefont{and}
  \bibinfo{author}{\bibfnamefont{D.}~\bibnamefont{Smith}},
  \bibinfo{journal}{Physical Review Letters} \textbf{\bibinfo{volume}{109}},
  \bibinfo{pages}{166605} (\bibinfo{year}{2012}).

\bibitem[{\citenamefont{Johnson et~al.}(2005)\citenamefont{Johnson, Petta,
  Taylor, Yacoby, Lukin, Marcus, Hanson, and Gossard}}]{johnson2005triplet}
\bibinfo{author}{\bibfnamefont{A.}~\bibnamefont{Johnson}},
  \bibinfo{author}{\bibfnamefont{J.}~\bibnamefont{Petta}},
  \bibinfo{author}{\bibfnamefont{J.}~\bibnamefont{Taylor}},
  \bibinfo{author}{\bibfnamefont{A.}~\bibnamefont{Yacoby}},
  \bibinfo{author}{\bibfnamefont{M.}~\bibnamefont{Lukin}},
  \bibinfo{author}{\bibfnamefont{C.}~\bibnamefont{Marcus}},
  \bibinfo{author}{\bibfnamefont{M.}~\bibnamefont{Hanson}}, \bibnamefont{and}
  \bibinfo{author}{\bibfnamefont{A.}~\bibnamefont{Gossard}},
  \bibinfo{journal}{Nature} \textbf{\bibinfo{volume}{435}},
  \bibinfo{pages}{925} (\bibinfo{year}{2005}).

\bibitem[{\citenamefont{Xu et~al.}(2008)\citenamefont{Xu, Sun, Berman, Steel,
  Bracker, Gammon, and Sham}}]{xu2008coherent}
\bibinfo{author}{\bibfnamefont{X.}~\bibnamefont{Xu}},
  \bibinfo{author}{\bibfnamefont{B.}~\bibnamefont{Sun}},
  \bibinfo{author}{\bibfnamefont{P.~R.} \bibnamefont{Berman}},
  \bibinfo{author}{\bibfnamefont{D.~G.} \bibnamefont{Steel}},
  \bibinfo{author}{\bibfnamefont{A.~S.} \bibnamefont{Bracker}},
  \bibinfo{author}{\bibfnamefont{D.}~\bibnamefont{Gammon}}, \bibnamefont{and}
  \bibinfo{author}{\bibfnamefont{L.}~\bibnamefont{Sham}},
  \bibinfo{journal}{Nature Physics} \textbf{\bibinfo{volume}{4}},
  \bibinfo{pages}{692} (\bibinfo{year}{2008}).

\bibitem[{\citenamefont{Press et~al.}(2010)\citenamefont{Press, De~Greve,
  McMahon, Ladd, Friess, Schneider, Kamp, H{\"o}fling, Forchel, and
  Yamamoto}}]{press2010ultrafast}
\bibinfo{author}{\bibfnamefont{D.}~\bibnamefont{Press}},
  \bibinfo{author}{\bibfnamefont{K.}~\bibnamefont{De~Greve}},
  \bibinfo{author}{\bibfnamefont{P.~L.} \bibnamefont{McMahon}},
  \bibinfo{author}{\bibfnamefont{T.~D.} \bibnamefont{Ladd}},
  \bibinfo{author}{\bibfnamefont{B.}~\bibnamefont{Friess}},
  \bibinfo{author}{\bibfnamefont{C.}~\bibnamefont{Schneider}},
  \bibinfo{author}{\bibfnamefont{M.}~\bibnamefont{Kamp}},
  \bibinfo{author}{\bibfnamefont{S.}~\bibnamefont{H{\"o}fling}},
  \bibinfo{author}{\bibfnamefont{A.}~\bibnamefont{Forchel}}, \bibnamefont{and}
  \bibinfo{author}{\bibfnamefont{Y.}~\bibnamefont{Yamamoto}},
  \bibinfo{journal}{Nature Photonics} \textbf{\bibinfo{volume}{4}},
  \bibinfo{pages}{367} (\bibinfo{year}{2010}).

\bibitem[{\citenamefont{Kuhlmann et~al.}(2013)\citenamefont{Kuhlmann, Houel,
  Ludwig, Greuter, Reuter, Wieck, Poggio, and Warburton}}]{kuhlmann2013charge}
\bibinfo{author}{\bibfnamefont{A.~V.} \bibnamefont{Kuhlmann}},
  \bibinfo{author}{\bibfnamefont{J.}~\bibnamefont{Houel}},
  \bibinfo{author}{\bibfnamefont{A.}~\bibnamefont{Ludwig}},
  \bibinfo{author}{\bibfnamefont{L.}~\bibnamefont{Greuter}},
  \bibinfo{author}{\bibfnamefont{D.}~\bibnamefont{Reuter}},
  \bibinfo{author}{\bibfnamefont{A.~D.} \bibnamefont{Wieck}},
  \bibinfo{author}{\bibfnamefont{M.}~\bibnamefont{Poggio}}, \bibnamefont{and}
  \bibinfo{author}{\bibfnamefont{R.~J.} \bibnamefont{Warburton}},
  \bibinfo{journal}{Nature Physics} \textbf{\bibinfo{volume}{9}},
  \bibinfo{pages}{570} (\bibinfo{year}{2013}).

\bibitem[{\citenamefont{Chekhovich et~al.}(2014)\citenamefont{Chekhovich,
  Hopkinson, Skolnick, and Tartakovskii}}]{chekhovich2014quadrupolar}
\bibinfo{author}{\bibfnamefont{E.}~\bibnamefont{Chekhovich}},
  \bibinfo{author}{\bibfnamefont{M.}~\bibnamefont{Hopkinson}},
  \bibinfo{author}{\bibfnamefont{M.}~\bibnamefont{Skolnick}}, \bibnamefont{and}
  \bibinfo{author}{\bibfnamefont{A.}~\bibnamefont{Tartakovskii}},
  \bibinfo{journal}{arXiv preprint arXiv:1403.1510}  (\bibinfo{year}{2014}).

\bibitem[{\citenamefont{Matthiesen et~al.}(2014)\citenamefont{Matthiesen,
  Stanley, Hugues, Clarke, and Atat{\"u}re}}]{matthiesen2014full}
\bibinfo{author}{\bibfnamefont{C.}~\bibnamefont{Matthiesen}},
  \bibinfo{author}{\bibfnamefont{M.~J.} \bibnamefont{Stanley}},
  \bibinfo{author}{\bibfnamefont{M.}~\bibnamefont{Hugues}},
  \bibinfo{author}{\bibfnamefont{E.}~\bibnamefont{Clarke}}, \bibnamefont{and}
  \bibinfo{author}{\bibfnamefont{M.}~\bibnamefont{Atat{\"u}re}},
  \bibinfo{journal}{Scientific Reports} \textbf{\bibinfo{volume}{4}},
  \bibinfo{pages}{4911} (\bibinfo{year}{2014}).

\bibitem[{\citenamefont{Hansom et~al.}(2014)\citenamefont{Hansom, Schulte,
  Le~Gall, Matthiesen, Clarke, Hugues, Taylor, and
  Atat{\"u}re}}]{hansom2014environment}
\bibinfo{author}{\bibfnamefont{J.}~\bibnamefont{Hansom}},
  \bibinfo{author}{\bibfnamefont{C.~H.~H.} \bibnamefont{Schulte}},
  \bibinfo{author}{\bibfnamefont{C.}~\bibnamefont{Le~Gall}},
  \bibinfo{author}{\bibfnamefont{C.}~\bibnamefont{Matthiesen}},
  \bibinfo{author}{\bibfnamefont{E.}~\bibnamefont{Clarke}},
  \bibinfo{author}{\bibfnamefont{M.}~\bibnamefont{Hugues}},
  \bibinfo{author}{\bibfnamefont{J.~M.} \bibnamefont{Taylor}},
  \bibnamefont{and}
  \bibinfo{author}{\bibfnamefont{M.}~\bibnamefont{Atat{\"u}re}},
  \bibinfo{journal}{Nature Physics} \textbf{\bibinfo{volume}{10}},
  \bibinfo{pages}{725} (\bibinfo{year}{2014}).

\bibitem[{\citenamefont{Matthiesen et~al.}(2012)\citenamefont{Matthiesen,
  Vamivakas, and Atat{\"u}re}}]{matthiesen2012subnatural}
\bibinfo{author}{\bibfnamefont{C.}~\bibnamefont{Matthiesen}},
  \bibinfo{author}{\bibfnamefont{A.~N.} \bibnamefont{Vamivakas}},
  \bibnamefont{and}
  \bibinfo{author}{\bibfnamefont{M.}~\bibnamefont{Atat{\"u}re}},
  \bibinfo{journal}{Physical Review Letters} \textbf{\bibinfo{volume}{108}},
  \bibinfo{pages}{093602} (\bibinfo{year}{2012}).

\bibitem[{\citenamefont{Magde et~al.}(1972)\citenamefont{Magde, Elson, and
  Webb}}]{magde1972thermodynamic}
\bibinfo{author}{\bibfnamefont{D.}~\bibnamefont{Magde}},
  \bibinfo{author}{\bibfnamefont{E.}~\bibnamefont{Elson}}, \bibnamefont{and}
  \bibinfo{author}{\bibfnamefont{W.~W.} \bibnamefont{Webb}},
  \bibinfo{journal}{Physical Review Letters} \textbf{\bibinfo{volume}{29}},
  \bibinfo{pages}{705} (\bibinfo{year}{1972}).

\bibitem[{\citenamefont{Krichevsky and
  Bonnet}(2002)}]{krichevsky2002fluorescence}
\bibinfo{author}{\bibfnamefont{O.}~\bibnamefont{Krichevsky}} \bibnamefont{and}
  \bibinfo{author}{\bibfnamefont{G.}~\bibnamefont{Bonnet}},
  \bibinfo{journal}{Reports on Progress in Physics}
  \textbf{\bibinfo{volume}{65}}, \bibinfo{pages}{251} (\bibinfo{year}{2002}).

\bibitem[{\citenamefont{Lippitz et~al.}(2005)\citenamefont{Lippitz, Kulzer, and
  Orrit}}]{lippitz2005statistical}
\bibinfo{author}{\bibfnamefont{M.}~\bibnamefont{Lippitz}},
  \bibinfo{author}{\bibfnamefont{F.}~\bibnamefont{Kulzer}}, \bibnamefont{and}
  \bibinfo{author}{\bibfnamefont{M.}~\bibnamefont{Orrit}},
  \bibinfo{journal}{ChemPhysChem} \textbf{\bibinfo{volume}{6}},
  \bibinfo{pages}{770} (\bibinfo{year}{2005}).

\bibitem[{\citenamefont{Machlup}(1954)}]{machlup1954noise}
\bibinfo{author}{\bibfnamefont{S.}~\bibnamefont{Machlup}},
  \bibinfo{journal}{Journal of Applied Physics} \textbf{\bibinfo{volume}{25}}
  (\bibinfo{year}{1954}).

\bibitem[{\citenamefont{Lai et~al.}(2006)\citenamefont{Lai, Maletinsky,
  Badolato, and Imamoglu}}]{lai2006knight}
\bibinfo{author}{\bibfnamefont{C.}~\bibnamefont{Lai}},
  \bibinfo{author}{\bibfnamefont{P.}~\bibnamefont{Maletinsky}},
  \bibinfo{author}{\bibfnamefont{A.}~\bibnamefont{Badolato}}, \bibnamefont{and}
  \bibinfo{author}{\bibfnamefont{A.}~\bibnamefont{Imamoglu}},
  \bibinfo{journal}{Physical Review Letters} \textbf{\bibinfo{volume}{96}},
  \bibinfo{pages}{167403} (\bibinfo{year}{2006}).

\bibitem[{\citenamefont{Welander et~al.}(2014)\citenamefont{Welander,
  Chekhovich, Tarttakovskii, and Burkard}}]{welander2014influence}
\bibinfo{author}{\bibfnamefont{E.}~\bibnamefont{Welander}},
  \bibinfo{author}{\bibfnamefont{E.}~\bibnamefont{Chekhovich}},
  \bibinfo{author}{\bibfnamefont{A.}~\bibnamefont{Tarttakovskii}},
  \bibnamefont{and} \bibinfo{author}{\bibfnamefont{G.}~\bibnamefont{Burkard}},
  \bibinfo{journal}{arXiv preprint arXiv:1405.1329}  (\bibinfo{year}{2014}).

\bibitem[{\citenamefont{Lu et~al.}(2010)\citenamefont{Lu, Zhao, Vamivakas,
  Matthiesen, F\"alt, Badolato, and Atat\"ure}}]{lu2010direct}
\bibinfo{author}{\bibfnamefont{C.-Y.} \bibnamefont{Lu}},
  \bibinfo{author}{\bibfnamefont{Y.}~\bibnamefont{Zhao}},
  \bibinfo{author}{\bibfnamefont{A.~N.} \bibnamefont{Vamivakas}},
  \bibinfo{author}{\bibfnamefont{C.}~\bibnamefont{Matthiesen}},
  \bibinfo{author}{\bibfnamefont{S.}~\bibnamefont{F\"alt}},
  \bibinfo{author}{\bibfnamefont{A.}~\bibnamefont{Badolato}}, \bibnamefont{and}
  \bibinfo{author}{\bibfnamefont{M.}~\bibnamefont{Atat\"ure}},
  \bibinfo{journal}{Phys. Rev. B} \textbf{\bibinfo{volume}{81}},
  \bibinfo{pages}{035332} (\bibinfo{year}{2010}).

\bibitem[{\citenamefont{Bechtold et~al.}(2014)\citenamefont{Bechtold, Rauch,
  Simmet, Ardelt, Regler, M{\"u}ller, and Finley}}]{bechtold2014hyperfine}
\bibinfo{author}{\bibfnamefont{A.}~\bibnamefont{Bechtold}},
  \bibinfo{author}{\bibfnamefont{D.}~\bibnamefont{Rauch}},
  \bibinfo{author}{\bibfnamefont{T.}~\bibnamefont{Simmet}},
  \bibinfo{author}{\bibfnamefont{P.}~\bibnamefont{Ardelt}},
  \bibinfo{author}{\bibfnamefont{A.}~\bibnamefont{Regler}},
  \bibinfo{author}{\bibfnamefont{K.}~\bibnamefont{M{\"u}ller}},
  \bibnamefont{and} \bibinfo{author}{\bibfnamefont{J.}~\bibnamefont{Finley}},
  \bibinfo{journal}{arXiv preprint arXiv:1410.4316v1}  (\bibinfo{year}{2014}).

\bibitem[{\citenamefont{Maletinsky et~al.}(2007)\citenamefont{Maletinsky,
  Badolato, and Imamoglu}}]{maletinsky2007dynamics}
\bibinfo{author}{\bibfnamefont{P.}~\bibnamefont{Maletinsky}},
  \bibinfo{author}{\bibfnamefont{A.}~\bibnamefont{Badolato}}, \bibnamefont{and}
  \bibinfo{author}{\bibfnamefont{A.}~\bibnamefont{Imamoglu}},
  \bibinfo{journal}{Physical Review Letters} \textbf{\bibinfo{volume}{99}},
  \bibinfo{pages}{056804} (\bibinfo{year}{2007}).

\bibitem[{\citenamefont{Latta et~al.}(2011)\citenamefont{Latta, Srivastava, and
  Imamo{\u{g}}lu}}]{latta2011hyperfine}
\bibinfo{author}{\bibfnamefont{C.}~\bibnamefont{Latta}},
  \bibinfo{author}{\bibfnamefont{A.}~\bibnamefont{Srivastava}},
  \bibnamefont{and}
  \bibinfo{author}{\bibfnamefont{A.}~\bibnamefont{Imamo{\u{g}}lu}},
  \bibinfo{journal}{Physical Review Letters} \textbf{\bibinfo{volume}{107}},
  \bibinfo{pages}{167401} (\bibinfo{year}{2011}).

\bibitem[{\citenamefont{Deng and Hu}(2008)}]{deng2008electron}
\bibinfo{author}{\bibfnamefont{C.}~\bibnamefont{Deng}} \bibnamefont{and}
  \bibinfo{author}{\bibfnamefont{X.}~\bibnamefont{Hu}},
  \bibinfo{journal}{Physical Review B} \textbf{\bibinfo{volume}{78}},
  \bibinfo{pages}{245301} (\bibinfo{year}{2008}).

\bibitem[{\citenamefont{Paget et~al.}(2008)\citenamefont{Paget, Amand, and
  Korb}}]{paget2008light}
\bibinfo{author}{\bibfnamefont{D.}~\bibnamefont{Paget}},
  \bibinfo{author}{\bibfnamefont{T.}~\bibnamefont{Amand}}, \bibnamefont{and}
  \bibinfo{author}{\bibfnamefont{J.-P.} \bibnamefont{Korb}},
  \bibinfo{journal}{Physical Review B} \textbf{\bibinfo{volume}{77}},
  \bibinfo{pages}{245201} (\bibinfo{year}{2008}).

\bibitem[{\citenamefont{Braun et~al.}(2005)\citenamefont{Braun, Marie, Lombez,
  Urbaszek, Amand, Renucci, Kalevich, Kavokin, Krebs, Voisin
  et~al.}}]{braun2005direct}
\bibinfo{author}{\bibfnamefont{P.-F.} \bibnamefont{Braun}},
  \bibinfo{author}{\bibfnamefont{X.}~\bibnamefont{Marie}},
  \bibinfo{author}{\bibfnamefont{L.}~\bibnamefont{Lombez}},
  \bibinfo{author}{\bibfnamefont{B.}~\bibnamefont{Urbaszek}},
  \bibinfo{author}{\bibfnamefont{T.}~\bibnamefont{Amand}},
  \bibinfo{author}{\bibfnamefont{P.}~\bibnamefont{Renucci}},
  \bibinfo{author}{\bibfnamefont{V.~K.} \bibnamefont{Kalevich}},
  \bibinfo{author}{\bibfnamefont{K.~V.} \bibnamefont{Kavokin}},
  \bibinfo{author}{\bibfnamefont{O.}~\bibnamefont{Krebs}},
  \bibinfo{author}{\bibfnamefont{P.}~\bibnamefont{Voisin}},
  \bibnamefont{et~al.}, \bibinfo{journal}{Phys. Rev. Lett.}
  \textbf{\bibinfo{volume}{94}}, \bibinfo{pages}{116601}
  (\bibinfo{year}{2005}).

\bibitem[{\citenamefont{Dreiser et~al.}(2008)\citenamefont{Dreiser, Atat\"ure,
  Galland, M\"uller, Badolato, and Imamoglu}}]{dreiser2008optical}
\bibinfo{author}{\bibfnamefont{J.}~\bibnamefont{Dreiser}},
  \bibinfo{author}{\bibfnamefont{M.}~\bibnamefont{Atat\"ure}},
  \bibinfo{author}{\bibfnamefont{C.}~\bibnamefont{Galland}},
  \bibinfo{author}{\bibfnamefont{T.}~\bibnamefont{M\"uller}},
  \bibinfo{author}{\bibfnamefont{A.}~\bibnamefont{Badolato}}, \bibnamefont{and}
  \bibinfo{author}{\bibfnamefont{A.}~\bibnamefont{Imamoglu}},
  \bibinfo{journal}{Physical Review B} \textbf{\bibinfo{volume}{77}},
  \bibinfo{pages}{075317} (\bibinfo{year}{2008}).

\bibitem[{\citenamefont{Houel et~al.}(2012)\citenamefont{Houel, Kuhlmann,
  Greuter, Xue, Poggio, Gerardot, Dalgarno, Badolato, Petroff, Ludwig
  et~al.}}]{houel2012probing}
\bibinfo{author}{\bibfnamefont{J.}~\bibnamefont{Houel}},
  \bibinfo{author}{\bibfnamefont{A.}~\bibnamefont{Kuhlmann}},
  \bibinfo{author}{\bibfnamefont{L.}~\bibnamefont{Greuter}},
  \bibinfo{author}{\bibfnamefont{F.}~\bibnamefont{Xue}},
  \bibinfo{author}{\bibfnamefont{M.}~\bibnamefont{Poggio}},
  \bibinfo{author}{\bibfnamefont{B.}~\bibnamefont{Gerardot}},
  \bibinfo{author}{\bibfnamefont{P.}~\bibnamefont{Dalgarno}},
  \bibinfo{author}{\bibfnamefont{A.}~\bibnamefont{Badolato}},
  \bibinfo{author}{\bibfnamefont{P.}~\bibnamefont{Petroff}},
  \bibinfo{author}{\bibfnamefont{A.}~\bibnamefont{Ludwig}},
  \bibnamefont{et~al.}, \bibinfo{journal}{Physical Review Letters}
  \textbf{\bibinfo{volume}{108}}, \bibinfo{pages}{107401}
  (\bibinfo{year}{2012}).

\bibitem[{\citenamefont{Nguyen et~al.}(2013)\citenamefont{Nguyen, Sallen,
  Abbarchi, Ferreira, Voisin, Roussignol, Cassabois, and
  Diederichs}}]{nguyen2013photoneutralization}
\bibinfo{author}{\bibfnamefont{H.~S.} \bibnamefont{Nguyen}},
  \bibinfo{author}{\bibfnamefont{G.}~\bibnamefont{Sallen}},
  \bibinfo{author}{\bibfnamefont{M.}~\bibnamefont{Abbarchi}},
  \bibinfo{author}{\bibfnamefont{R.}~\bibnamefont{Ferreira}},
  \bibinfo{author}{\bibfnamefont{C.}~\bibnamefont{Voisin}},
  \bibinfo{author}{\bibfnamefont{P.}~\bibnamefont{Roussignol}},
  \bibinfo{author}{\bibfnamefont{G.}~\bibnamefont{Cassabois}},
  \bibnamefont{and}
  \bibinfo{author}{\bibfnamefont{C.}~\bibnamefont{Diederichs}},
  \bibinfo{journal}{Physical Review B} \textbf{\bibinfo{volume}{87}},
  \bibinfo{pages}{115305} (\bibinfo{year}{2013}).

\bibitem[{\citenamefont{Davan{\c{c}}o et~al.}(2014)\citenamefont{Davan{\c{c}}o,
  Hellberg, Ates, Badolato, and Srinivasan}}]{davancco2014multiple}
\bibinfo{author}{\bibfnamefont{M.}~\bibnamefont{Davan{\c{c}}o}},
  \bibinfo{author}{\bibfnamefont{C.~S.} \bibnamefont{Hellberg}},
  \bibinfo{author}{\bibfnamefont{S.}~\bibnamefont{Ates}},
  \bibinfo{author}{\bibfnamefont{A.}~\bibnamefont{Badolato}}, \bibnamefont{and}
  \bibinfo{author}{\bibfnamefont{K.}~\bibnamefont{Srinivasan}},
  \bibinfo{journal}{Physical Review B} \textbf{\bibinfo{volume}{89}},
  \bibinfo{pages}{161303} (\bibinfo{year}{2014}).

\bibitem[{\citenamefont{Fallahi et~al.}(2010)\citenamefont{Fallahi, Y{\i}lmaz,
  and Imamo{\u{g}}lu}}]{fallahi2010measurement}
\bibinfo{author}{\bibfnamefont{P.}~\bibnamefont{Fallahi}},
  \bibinfo{author}{\bibfnamefont{S.}~\bibnamefont{Y{\i}lmaz}},
  \bibnamefont{and}
  \bibinfo{author}{\bibfnamefont{A.}~\bibnamefont{Imamo{\u{g}}lu}},
  \bibinfo{journal}{Physical Review Letters} \textbf{\bibinfo{volume}{105}},
  \bibinfo{pages}{257402} (\bibinfo{year}{2010}).

\bibitem[{\citenamefont{Chekhovich et~al.}(2011)\citenamefont{Chekhovich,
  Krysa, Skolnick, and Tartakovskii}}]{chekhovich2011direct}
\bibinfo{author}{\bibfnamefont{E.}~\bibnamefont{Chekhovich}},
  \bibinfo{author}{\bibfnamefont{A.}~\bibnamefont{Krysa}},
  \bibinfo{author}{\bibfnamefont{M.}~\bibnamefont{Skolnick}}, \bibnamefont{and}
  \bibinfo{author}{\bibfnamefont{A.}~\bibnamefont{Tartakovskii}},
  \bibinfo{journal}{Physical Review Letters} \textbf{\bibinfo{volume}{106}},
  \bibinfo{pages}{027402} (\bibinfo{year}{2011}).

\bibitem[{\citenamefont{Hooge and Bobbert}(1997)}]{hooge1997correlation}
\bibinfo{author}{\bibfnamefont{F.}~\bibnamefont{Hooge}} \bibnamefont{and}
  \bibinfo{author}{\bibfnamefont{P.}~\bibnamefont{Bobbert}},
  \bibinfo{journal}{Physica B: Condensed Matter}
  \textbf{\bibinfo{volume}{239}}, \bibinfo{pages}{223} (\bibinfo{year}{1997}).

\end{thebibliography}

\end{document}